\documentclass[10pt]{revtex4}
\usepackage{color}
\usepackage{bbm}
\usepackage{epsfig,amssymb,amsbsy,amsmath}
\usepackage[version=4]{mhchem}

\newtheorem{prop}{Proposition}

\begin{document}
%\tightenlines \draft

\title{Generalized Bell inequalities and frustrated spin systems
}
\author{Heinz-J\"urgen Schmidt}

\address{Universit\"at Osnabr\"uck, Fachbereich Mathematik, Informatik und Physik,
Barbarastr.~7, 49069 Osnabr\"uck, Germany}

\begin{abstract}
We find a close correspondence between generalized Bell
inequalities (GBI's) of a special kind and certain frustrated spin
systems. For example, the Clauser-Horn-Shimony-Holt inequality
corresponds to the frustrated square with the signs $+++-$ for the
nearest neighbor interaction between the spins. Similarly, the
Pearle-Braunstein-Cave inequality corresponds to a frustrated even
ring with the corresponding signs $+\ldots +-$. Upon this
correspondence, the violation of such inequalities by the
entangled singlet state in quantum mechanics is equivalent to the
spin system possessing a classical coplanar ground state, the
energy of which is lower than the Ising ground state's energy. We
propose a scheme which generates new inequalities and give further
examples, the frustrated hexagon with additional diagonal bonds
and the frustrated hypercubes in $n=3,4,5$ dimensions.
Surprisingly, the hypercube in $n=4$ dimensions yields an
inequality which is \emph{not} violated by the singlet state. We
extend the correspondence to other entangled states and XXZ-models
of spin systems.\\

\end{abstract}
\maketitle

\widetext
%%%%%%%%%%%%%%%%%%%%%%%%%%%%%%%%%%%%%%%%%%%%%%%%%%%%%%%%%%%%%%%%%%%%%%%%%%%%%%%%%%%%%%%%%%%%%%%%%
%%%%%%%%%%%%%%%%%%%%%%%%%%%%%%%%%%%%%%%%%%%%%%%%%%%%%%%%%%%%%%%%%%%%%%%%%%%%%%%%%%%%%%%%%%%%%%%%%
\section{Introduction\label{sec:I}}
%%%%%%%%%%%%%%%%%%%%%%%%%%%%%%%%%%%%%%%%%%%%%%%%%%%%%%%%%%%%%%%%%%%%%%%%%%%%%%%%%%%%%%%%%%%%%%%%%
%%%%%%%%%%%%%%%%%%%%%%%%%%%%%%%%%%%%%%%%%%%%%%%%%%%%%%%%%%%%%%%%%%%%%%%%%%%%%%%%%%%%%%%%%%%%%%%%%
Bell's inequality, published more than six decades ago, has not
ceased to invoke keen interest in the physics community. The title
of the seminal paper of J.~Bell \cite{Bell:1964} refers to the
famous article of A.~Einstein, B.~Podolski, and N.~Rosen
\cite{EPR:1935} (EPR) who concluded that, according to their
criteria, quantum theory (QT) is incomplete. Bell proved that the
assumptions of EPR lead to an inequality for measurable
correlations of spin measurements for two particles which is
violated in QT, and, as later work showed, also in experiments.
Thus certain kinds of hidden variable theories are empirically ruled out.
Of course, the exact relation between Bell's assumptions and those of
EPR has to be carefully examined. According to an analysis of
L.~E.~Ballentine and J.~P.~Jarrett \cite{BJ:1987} the assumptions
of Bell, simple locality and predictive completeness, are even
weaker than the EPR assumptions. Hence, in the words of these
authors, the \emph{incompleteness} of QT \emph{is, in some sense, a
property of nature}
\cite{BJ:1987}.\\

There have been many proposals to generalize Bell's inequality. An
important generalization is the Clauser-Horn-Shimony-Holt
inequality \cite{CHSH:1969} (CHSH) which is about a linear
combination of the correlations of two pairs of measurements. A
generalization to $n$ pairs of measurements has been considered by
Pearle \cite{P:1970} and later investigated by Braunstein and Cave
\cite{BC:1990}. Other work generalized Bell's inequality to an
arbitrary number of measurements \cite{GS:1979} or to more than
two particles \cite{AP:1993}. See the textbook of A.~Peres
\cite{AP:1993}
and literature quoted there for more details.
Recently, remarkable graph-theoretic and algebraic approaches have been
developed to provide a framework for analyzing quantum nonlocality in large one-dimensional systems
\cite{EHAT:2024}  \cite{Hetal:2026}.
\\

In this article we will point out a close correspondence between
possible generalized Bell inequalities (GBI's) and certain
frustrated \emph{classical} spin systems $\Sigma_N^{cl}$.
These terms will be explained in more detail below.
It is important to distinguish the spin system $\Sigma_N^{cl}$ from
the \emph{quantum} spin system $\Sigma_2^q$ on which the
EPR measurements are performed.
The number $N$ of spins in $\Sigma_N^{cl}$ corresponds to
the number of measurements considered in the context of GBI's.
As a by-product of the correspondence we will obtain a procedure
to generate new GBI's including a test whether these inequalities are
violated in QT.\\

This paper is organized as follows: In Section \ref{sec:G} we
explain the basic idea of the correspondence using the example of
the CHSH inequality. Then we will give the general definitions for
the spin systems which give rise to a correspondence with GBI's.
The GBI is violated by the singlet state if and only if the
corresponding Heisenberg spin system has a classical ground state
with a lower energy than the corresponding Ising ground state.
These systems thus have necessarily non-collinear ground state
configurations, i.~e.~coplanar or $3$-dimensional ones, although
in all examples considered in this article it is not necessary to
consider $3$-dimensional ground states. In Section \ref{sec:C} we
present some methods to calculate classical ground states and
apply these to the construction procedure for GBI's in Section \ref{sec:B}.
Sometimes new GBI's can be found by merging spins;
also the original Bell inequality adapted to
experiments with linearly polarized pairs of photons can
be obtained from the Bell square by merging two spins
and thus obtaining a frustrated spin triangle, see Subsection \ref{sec:MSV},
where this procedure is generalized.

Section \ref{sec:E} is devoted to a couple of
examples, including the frustrated $2n$-ring leading to the
Pearle-Braunstein-Cave inequality, the frustrated hexagon and the
frustrated hypercubes $H_n$. In all these examples it is possible
to analytically calculate classical Heisenberg ground states and
Ising ground states and to compare their ground state energies.
With the exception of $H_4$ the Ising ground state energy is
higher and hence we obtain GBI's violated in QT by the singlet
state. More general entangled states are considered in Section
\ref{sec:S} and are shown to lead to XXZ-models of spin systems.
We close with a conclusion in Section \ref{sec:CC}. The
correspondence between GBI's and frustrated spin systems is
summarized in table \ref{tab1}.
\\

%%%%%%%%%%%%%%%%%%%%%%%%%%%%%%%%%%%%%%%%%%%%%%%%%%%%%%%%%%%%%%%%%%%%%%%%%%%%%%%%%%%%%%%%%%%%%%%%%
%%%%%%%%%%%%%%%%%%%%%%%%%%%%%%%%%%%%%%%%%%%%%%%%%%%%%%%%%%%%%%%%%%%%%%%%%%%%%%%%%%%%%%%%%%%%%%%%%
\section{Inequalities and spin systems\label{sec:G}}
%%%%%%%%%%%%%%%%%%%%%%%%%%%%%%%%%%%%%%%%%%%%%%%%%%%%%%%%%%%%%%%%%%%%%%%%%%%%%%%%%%%%%%%%%%%%%%%%%
\subsection{Bell inequalities}\label{sec:BI}
%%%%%%%%%%%%%%%%%%%%%%%%%%%%%%%%%%%%%%%%%%%%%%%%%%%%%%%%%%%%%%%%%%%%%%%%%%%%%%%%%%%%%%%%%%%%%%%%%

In order to motivate the correspondence between generalized Bell
inequalities (GBI's) and frustrated spin systems
we consider the CHSH inequality \cite{CHSH:1969},
following \cite{AP:1993}.\\
Let $a,b,c,d$ be four numbers which assume only
the values $\pm 1$. Then either $a+c=0$ or $a-c=0$ and hence
$(a+c)b+(a-c)d$ assumes only the values $\pm 2$,
which yields the inequality
\begin{equation}\label{G1}
-2\le ab+ad+cb-cd \le 2
\;.
\end{equation}
Imagine that the numbers $a_i,b_i,c_i,d_i,\; i=1,\ldots,N$ are the
outcomes of four $N$-times repeated experiments and consider the
mean values of the above products, i.~e.~the correlation,
\begin{equation}\label{G2}
\langle a b \rangle \equiv \frac{1}{N}\sum_{i=1}^N
a_i\, b_i=\langle b a \rangle, \;\text{ etc. }
\;.
\end{equation}
Then the inequality (\ref{G1}) also holds for the mean values, i.~e.
\begin{equation}\label{G3}
-2\le \langle ab\rangle +\langle ad\rangle +
\langle cb\rangle -\langle cd\rangle  \le 2
\;.
\end{equation}
This is an equivalent form of the CHSH inequality \cite{CHSH:1969}
which belongs to the family of generalized Bell inequalities GBI.
It bounds the classical correlations between four different
measurements. We have formulated it
without the assumption that the mean values converge against some
expectation values for $N\rightarrow\infty$, following
\cite{AP:1993} since it seems to be more appropriate to consider
this as part of the postulates which connect the
CHSH inequality
to measurements or to QT. These postulates are known to lead to a
contradiction.
The CHSH inequality alone is a mathematically valid
statement about $N\times 4$-matrices with entries $\pm 1$.\\

%===================    figure   =================================

\begin{figure}
\begin{center}
\includegraphics[width=150mm]{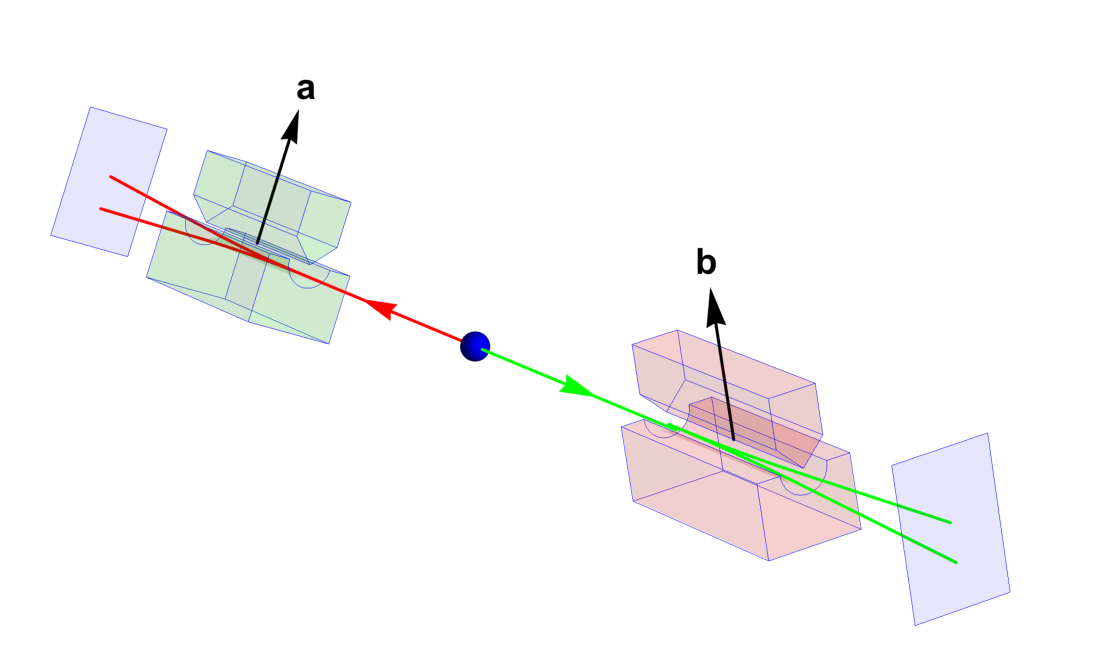}
\caption{\label{FIGEPR}
 A schematic sketch of a combined spin measurement in the direction
of $\vec{a}$ (performed by Alice) and $\vec{b}$ (performed by Bob) using Stern-Gerlach magnets.
}
\end{center}
\end{figure}

Now consider in QT a pair of particles with spin $s=\frac{1}{2}$
in its entangled singlet $(S=0)$ spin state
\begin{equation}\label{G4}
\phi=\frac{1}{\sqrt{2}}\left( |\uparrow\downarrow\,\rangle -
|\downarrow\uparrow\,\rangle\right) \;.
\end{equation}
Further consider measurements of the single particle spin in
direction of the unit vectors
$\vec{a},\vec{b},\vec{c},\vec{d}$.
In principle, these unit vectors can be chosen arbitrarily.
However, in all applications they will be coplanar,
that is, orthogonal to some common axis.
The corresponding observables are
represented by the Hermitean operators
\begin{equation}\label{G5}
A=\vec{a}\cdot \vec{\sigma} = a_x \sigma_x +
a_y \sigma_y +a_z \sigma_z
\;,
\end{equation}
where $\sigma_x,\sigma_y,\sigma_z$ are the Pauli matrices and the
observables $B,C,D$ for the other directions
$\vec{b},\vec{c},\vec{d}$ are analogously defined. It is possible
to combine any two of these measurements and to measure, say, $A$
at the left-hand particle and $B$ at the other one,
see Figure \ref{FIGEPR}.
The two experimenters doing these measurements are traditionally called
``Alice" and ``Bob". According to the rules of QT, the combined
measurement is represented by the tensor product operator
$A\otimes B$. If $a_i, b_i$ are the outcomes of $N$ repetitions of
this combined measurements, the mean values $\langle a b \rangle$
according to (\ref{G2}) converge towards the expectation value,
which can be calculated by QT and depends on the state of the
system. For the singlet state (\ref{G4}) the expectation value
turns out to be
\begin{equation}\label{G6}
\langle A B \rangle \equiv \langle \phi| A \otimes B
|\phi\rangle=-\vec{a}\cdot\vec{b}
\;,
\end{equation}
and analogously for $\langle A D \rangle,\langle C B \rangle$
and $\langle C D \rangle$. Hence the correlation term in
the CHSH inequality (\ref{G3}) has the quantum theoretical
counterpart
\begin{equation}\label{G7}
\langle A B \rangle +\langle A D \rangle +\langle C B \rangle -
\langle C D \rangle
=
-\vec{a}\cdot\vec{b} -\vec{a}\cdot\vec{d}-\vec{c}\cdot\vec{b}+
\vec{c}\cdot\vec{d}
\;.
\end{equation}
It is easily seen that the extreme values of (\ref{G7}) exceed
the bounds of the CHSH inequality. The possible values
of (\ref{G7}) are symmetric with respect to $0$, since the
substitution
$\vec{a}\mapsto -\vec{a}, \vec{c}\mapsto -\vec{c}$ changes
the overall sign in (\ref{G7}).
Writing (\ref{G7}) in the form $-\vec{a}\cdot(\vec{b} +\vec{d})-
\vec{c}\cdot(\vec{b}-\vec{d})$
it is obvious that each term is minimal for the choice
$\vec{a}\, |\,|\, (\vec{b} +\vec{d})$ and $\vec{c} \,
|\,|\,(\vec{b}-\vec{d})$. Moreover,
$|\vec{b} +\vec{d}|+|\vec{b} -\vec{d}|$ is maximal for
$\sphericalangle(\vec{b},\vec{d})=90^\circ$.
This is equivalent to the statement: The square
has the maximal circumference among the rectangles
with fixed length of their diagonals. Hence
\begin{equation}\label{G8a}
-2\sqrt{2} \le \vec{a}\cdot\vec{b} +\vec{a}\cdot\vec{d}+
\vec{c}\cdot\vec{b}-\vec{c}\cdot\vec{d}\le 2\sqrt{2}
\;,
\end{equation}
or
\begin{equation}\label{G8b}
-2\sqrt{2} \le
\langle A B \rangle +\langle A D \rangle +\langle C B \rangle -
\langle C D \rangle
\le 2\sqrt{2}
\;,
\end{equation}
and the lower bound is assumed for any
planar spin
configuration with
$\sphericalangle(\vec{b},\vec{d})=90^\circ,
\sphericalangle(\vec{c},\vec{d})=45^\circ,
\sphericalangle(\vec{a},\vec{c})=90^\circ$, see figure \ref{fig04a}.
The upper bound is assumed similarly.
Here and in what follows we always count the
angles $\sphericalangle(\vec{a},\vec{b})$ between two unit vectors
counter-clockwise, beginning with $\vec{a}$.
Equation (\ref{G8b}) expresses the bounds for quantum correlations
between four possible measurements. Since the bounds are attained
the CHSH inequality (\ref{G3}) is violated in QT for suitable
measurements and entangled states.
\\

The violation of the CHSH and similar inequalities,
which are rigorously proven theorems, can only be understood
in the sense that some
of the assumptions leading to (\ref{G3}) must
not hold in QT. Indeed, if $[A,C]\neq 0$ and $[B,D]\neq 0$,
only one of the four possible combinations
of measurements $AB, AD, CB, CD$ can be performed as a joint
measurement and hence only two of the four numbers
$a, b, c, d$ can be actually measured in one single experiment.
If QT is right, it is thus not possible, by whatever means,
to predict the missing two numbers in a consistent way,
i.~e.~in such a way that the mean values of all correlation
measurements
(actual and hypothetical ones) approach the mean values of the actual
correlation measurements alone.
In the words of A.~Peres \cite{AP:1978}:
\emph{ Unperformed experiments have no results}.\\

%-------------------------table-------------------------------
\begin{table}
\caption{\label{tab2}Correspondence between GBI's and
classical spin systems.}
\vspace{5mm} \item[]\begin{tabular}{ll} \hline
EPR experiment & Classical spin system\\
\hline
$N$ possible measurements & $N$ spins \\
$2$-particle states & bi-linear Hamiltonian \\
isotropic singlet state & isotropic Heisenberg Hamiltonian $H(\mathbf{s})$\\
general entangled state & XXZ-Hamiltonian, Section \ref{sec:S} \\
$2$ experimenters (Alice and Bob) & AB-systems\\
GBI & inequality for Ising states $s$ \\
$|\sum_{\mu\in{\mathcal A}}\sum_{\nu\in{\mathcal B}}J_{\mu\nu}
\langle s_\mu s_\nu \rangle|\le E_0^{(I)}$
&
$|H^{(I)}(s)|\le E_0^{(I)}$\\
Violation of the GBI for quantum correlations
&
non-collinear ground state state $\mathbf{s}$ \\
$\sum_{\mu\in{\mathcal A}}\sum_{\nu\in{\mathcal B}}J_{\mu\nu}
\vec{s}_\mu \cdot \vec{s}_\nu = - E_0$
&
$H(\mathbf{s})=- E_0$\\
such that $E_0^{(I)}< E_0$
&
such that $E_0^{(I)}< E_0$\\
\hline\hline
\end{tabular}
\end{table}
%-------------------------table-------------------------------

%===================    figure   =================================

\begin{figure}
\begin{center}
\includegraphics[width=150mm]{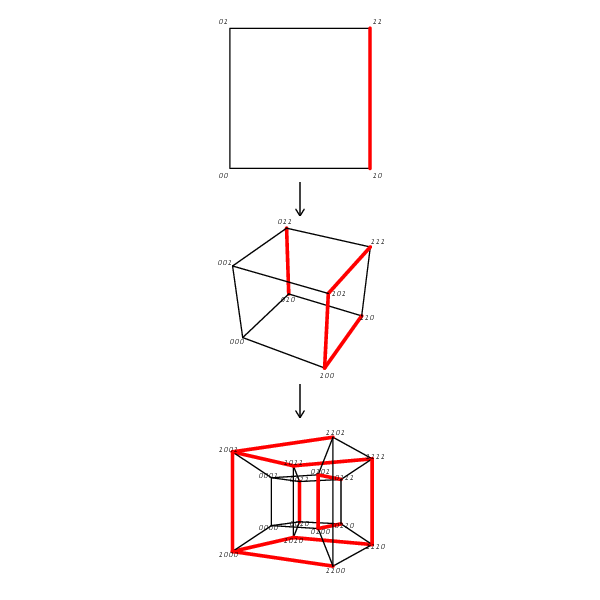}
\caption{\label{fig03}A sequence of frustrated hypercubes $H_n,\;n=2,3,4$ generated by a recursive procedure,
see Subsection \ref{sec:E3}. The negative bonds are indicated by thick red lines.
}
\end{center}
\end{figure}

%===================    figure   =================================
\begin{figure}[ht]
\begin{center}
\includegraphics[width=60mm]{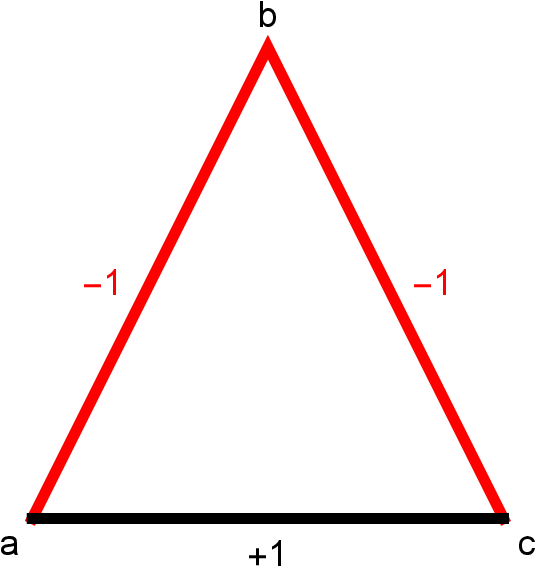}
\caption{\label{FIGT}Sketch of a frustrated spin triangle corresponding to the
Bell inequality (\ref{G3c}), see Subsection \ref{sec:MSV}.
}
\end{center}
\end{figure}
%===================    figur

\subsection{Correspondence to frustrated spin systems}\label{sec:CFS}

Next we introduce the correspondence to spin systems, see
table \ref{tab2}.
The basic idea is to re-interpret (\ref{G8a}) as a statement
about the energy
of a classical Heisenberg spin system $\Sigma$ of four spins,
which are represented by the unit vectors
$\vec{a},\vec{b},\vec{c},\vec{d}$. Of course, $\Sigma$ has nothing
to do with the quantum two-spin system for which the EPR type of
measurements are performed. The different signs in (\ref{G8a})
reflect the coupling between the four spins: $\Sigma$ can be
visualized as a square with three anti-ferromagnetic bonds
($J=+1$) and one ferromagnetic bond ($J=-1$), see (\ref{E31}) and
figure \ref{fig03}. Each spin configuration
$\vec{a},\vec{b},\vec{c},\vec{d}$ realizing the lower bound of
(\ref{G8a}) is thus a classical ground state for the
(dimensionless) Hamiltonian
\begin{equation}\label{HBell}
 H(\vec{a},\vec{b},\vec{c},\vec{d}) = -\vec{a}\cdot\vec{b} -\vec{a}\cdot\vec{d}-\vec{c}\cdot\vec{b}+
\vec{c}\cdot\vec{d}
\;.
\end{equation}

In this paper we will always denote the ground state energy of a spin system
by $-E_0$ in order to avoid inequalities of the form $E_0\le \ldots \le -E_0$.\\
The ground states do not minimize each term in (\ref{G8a})
separately. Indeed, each term in (\ref{G8a}) has the form of a spin dimer
with a classical ground state energy of $-1$, irrespective of the sign of the dimer term.
Adding these four ground state energies would result in
$-E_0=-4$ (the same as in the case where all four signs in
(\ref{G8a}) are $+1$). Actually, the ground state energy of (\ref{G8a})
is only $-E_0=-2\,\sqrt{2}$ since one cannot minimize each term  of (\ref{G8a})
separately without regard to the other terms.\\
Spin systems with this property,
namely where not each term in the Heisenberg
(or Ising) Hamiltonian can be minimized simultaneously by a classical ground state
will be called ``frustrated" throughout this article.
Note that it is \textit{not} excluded that the ground state of a frustrated spin system
may be attained by a collinear (Ising) state; see the frustrated hypercube $H_4$
in Subsection \ref{sec:E3} as an example.
Sometimes this notion
is used in the literature with slightly different meanings.
For the theory of frustrated spin systems, see, for example, \cite{SRFB:2004},
or, for an introduction, \cite{MR:2006}. One of the simplest
examples is the anti-ferromagnetic (AF) spin triangle where the
local states $\uparrow\downarrow$ minimize one term in the
Heisenberg Hamiltonian but cannot be extended to a global ground
state. The AF triangle has coplanar ground states with angles of
$120^\circ$ between the spin vectors and is said to be
``geometrically frustrated". The frustrated square (\ref{G8a}), or
``Bell square", as we will call it,
is an example of ``non-geometric frustration".
A square with bonds of the same sign is not frustrated; the frustration
of the Bell square is only due to the pattern of ferromagnetic and antiferromagnetic bonds.
\\

Alternatively, the ground state of the Bell square can be calculated
by the methods provided in \cite{FSJ:2026}, see Section \ref{sec:E1}.
For a spin ring that consists entirely of ferromagnetic bonds $J_{\mu,\mu+1}=-1$
except for a single antiferromagnetic bond $J_{01}=1$
we may anticipate the result that the ground state energy will be
\begin{equation}\label{E0a}
  -E_0 = - N \, \cos \frac{\pi}{N}
  \;.
\end{equation}
For the Bell square with $N=4$ it follows that $-E_0=-2\sqrt{2}$.\\

The Bell square is an ``AB-system", that is, the spins can
be divided into two sub-lattices ${\mathcal A}$ and
${\mathcal B}$ such that only ${\mathcal A}$-spins interact with
${\mathcal B}$-spins.
(The nomenclature is again reminiscent of ``Alice" and ``Bob").
A system with
\textit{red}{
The notion of AB-systems has to be distinguished
from the similar notion of ``bipartite systems", where there is only
non-positive interaction \emph{within} the sub-lattices
and non-negative interaction \emph{between} the sub-lattices.
Bipartite systems  have ground states of the form that, say,
all ${\mathcal A}$-spins are down and all ${\mathcal B}$-spins are up.
}
The Bell square is an AB-system, but not bi-partite.
This could be taken as a preliminary definition of
``non-geometric frustration". However,
we will not attempt to define ``geometrical frustration" and
``non-geometrical frustration"  more precisely and
will rather use these notions in a somewhat intuitive and vague sense.\\

Now we can also reinterpret the CHSH inequality, or, rather, its
precursor (\ref{G1}), as a statement about the Bell square spin
system: It simply says that its energy according to the Ising
model is bounded by $\pm 2$. In the Ising model, the individual
spin is not represented by a $3$-dimensional unit vector, but, so
to speak, by an $1$-dimensional unit vector $\uparrow$ or
$\downarrow$ , or, equivalently, by numbers $\pm 1$. The
bi-partite systems mentioned above have ground states of the form
$\uparrow,\; \downarrow$, i.~e.~collinear or Ising ground states.
But the Bell square Heisenberg spin system has a coplanar ground
state with a ground state energy $-E_0=-2\sqrt{2}$ below its Ising
model ground state energy $-E_0^{(I)}=-2$. This is the violation
of the CHSH inequality in QT,
translated into the language of spin systems.

\subsection{Construction of generalized Bell inequalities}\label{sec:CGBI}

We will try to exploit the described correspondence in order to
construct GBI's violated by the entangled singlet state in QT by considering
the corresponding non-geometrically frustrated
spin systems with only coplanar or $3$-dimensional ground states.
To this end we need some more general notation
which will be adapted to the described correspondence.\\

The state of the spin system $\Sigma$
is described by $N$ unit vectors $\vec{s}_\mu,\;\mu=1,\ldots,N$.
Moreover, the set of spins is divided into two
disjoint subsets, $\{1,\ldots,N\}={\mathcal A}\,
\stackrel{.}{\cup}\,{\mathcal B}$, such that the
Hamiltonian of $\Sigma$
can be written in the form
\begin{equation}\label{G10}
H=\sum_{\mu\in{\mathcal A}} \sum_{\nu\in{\mathcal B}}
J_{\mu\nu} \vec{s}_\mu\cdot\vec{s}_\nu
\;,
\end{equation}
where the $J_{\mu\nu}$ are real coupling coefficients,
some of which may vanish.
The minimum of the Hamiltonian (\ref{G10}) is called
the ground state energy $-E_0$.\\
The corresponding Ising model has states described by
numbers $s_\mu=\pm 1,\;\mu=1,\ldots,N$ and an Ising
Hamiltonian
\begin{equation}\label{G10a}
H^{(I)}=\sum_{\mu\in{\mathcal A}}
\sum_{\nu\in{\mathcal B}}J_{\mu\nu} s_\mu s_\nu
\end{equation}
with the minimum $-E_0^{(I)}$, the Ising ground state energy.
In both cases, the values of the Hamiltonian
change their sign under the spin flip transformation
$\vec{s}_\mu\mapsto - \vec{s}_\mu$, resp. $s_\mu\mapsto -s_\mu$,
$\mu\in{\mathcal A}$.\\

In the context of the EPR situation, the
$s_\mu=\pm 1,\,\mu=1,\ldots, N$
are the possible outcomes of one
experiment and the GBI assumes the form
\begin{equation}\label{G11}
-E_0^{(I)}\le \sum_{\mu\in{\mathcal A}}
\sum_{\nu\in{\mathcal B}}J_{\mu\nu}
\langle s_\mu s_\nu \rangle \le E_0^{(I)}
\;.
\end{equation}
The unit vectors $\vec{s}_\mu,\;\mu\in{\mathcal A},$
describe the directions of spin measurements
at, say, the left-hand particle (done by Alice),
and analogously
$\vec{s}_\mu,\;\mu\in{\mathcal B},$ for measurements
at the right-hand particle (done by Bob).
The quantum theoretical counter-part of (\ref{G11})
is the inequality
\begin{equation}\label{G12}
-E_0\le - \sum_{\mu\in{\mathcal A}}
\sum_{\nu\in{\mathcal B}} J_{\mu\nu}
\langle A_\mu B_\nu \rangle
=
\sum_{\mu\in{\mathcal A}} \sum_{\nu\in{\mathcal B}}
J_{\mu\nu} \vec{s}_\mu\cdot\vec{s}_\nu \le E_0
\;,
\end{equation}
where all correlations are calculated in the
singlet state (\ref{G4}). The GBI (\ref{G11})
is hence violated in QT if and only if $E_0^{(I)} < E_0$.
In all applications, the ground state of the spin system $\Sigma$
will be collinear or coplanar. For this reason,
it is assumed that the $N$ unit vectors $\vec{s}_\mu,\;\mu\in{\mathcal A}\cup {\mathcal B}$,
which describe the directions of the spin measurements intended to detect a violation of the GBI,
will be coplanar. However, there is no known general theorem that
rules out three-dimensional ground states for the AB systems considered here.

%%%%%%%%%%%%%%%%%%%%%%%%%%%%%%%%%%%%%%%%%%%%%%%%%%%%%%%%%%%%%%%%%%%%%%%%%%%%%%%%%%%%%%%%%%%%%%%%%
%%%%%%%%%%%%%%%%%%%%%%%%%%%%%%%%%%%%%%%%%%%%%%%%%%%%%%%%%%%%%%%%%%%%%%%%%%%%%%%%%%%%%%%%%%%%%%%%%
\section{Calculation of the ground state\label{sec:C}}
%%%%%%%%%%%%%%%%%%%%%%%%%%%%%%%%%%%%%%%%%%%%%%%%%%%%%%%%%%%%%%%%%%%%%%%%%%%%%%%%%%%%%%%%%%%%%%%%%
%%%%%%%%%%%%%%%%%%%%%%%%%%%%%%%%%%%%%%%%%%%%%%%%%%%%%%%%%%%%%%%%%%%%%%%%%%%%%%%%%%%%%%%%%%%%%%%%%

For the calculation of the classical ground state
of a spin system $\Sigma$ there exists no straightforward
method. However, sometimes the following considerations are useful.
For more details see \cite{SM:2003},
\cite{S:2017} and \cite{SR:2022}.
We write the coupling constants $J_{\mu\nu}$ as the entries
of a symmetric $N\times N$-matrix $\mathbb{J}$.
Let $j_{\text{\scriptsize min}}$ denote its lowest eigenvalue.
Then the Rayleigh-Ritz variation principle yields
\begin{equation}\label{C1}
2 H = \sum_{\mu,\nu=1}^N J_{\mu\nu}
\vec{s}_\mu\cdot\vec{s}_\nu \ge
j_{\text{\scriptsize min}}\sum_{\mu=1}^N
(\vec{s}_\mu)^2 = N j_{\text{\scriptsize min}}
\;,
\end{equation}
whence
\begin{equation}\label{C1a}
\frac{1}{2} N j_{\text{\scriptsize min}}\le -E_0
\;.
\end{equation}
In general, this is only a lower bound and we may have
$\frac{1}{2} N j_{\text{\scriptsize min}}< -E_0$.
However, if $(s_\mu^{(i)})_{\mu=1,\ldots,N},\; i=1,2$ are
two linearly independent eigenvectors of $\mathbb{J}$
with eigenvalue $j_{\text{\scriptsize min}}$ such that
$(s_\mu^{(1)})^2 + (s_\mu^{(2)})^2 =1$ for all
$\mu=1,\ldots,N$ then the inequality (\ref{C1}) shows that we
have found a coplanar ground state
\begin{equation}\label{C2}
\vec{s}_\mu =\left(\begin{array}{c}s_\mu^{(1)}\\
s_\mu^{(2)}\end{array}\right),\;
\mu=1,\ldots,N
\end{equation}
with ground state energy
$-E_0=\frac{1}{2} N j_{\text{\scriptsize min}}$.
Analogously we can argue for
three linearly independent eigenvectors which
yield a $3$-dimensional ground state of $H$.
%Further information on the theory of ground states
%of classical spin systems can be found
%in \cite{SR:2022} and the references cited therein.
\\

To find the Ising ground state $(s_\mu)_{\mu=1,\ldots,N}$, the
simplest method would be to check all $2^N$ Ising spin
configurations. Since we can choose, say, $s_1=1$ without loss of
generality, it would suffice to check $2^{N-1}$ states. But also
this can be a forbidding large number if $N$ is not too small.
Assume, for example, that we have $N=32$ spins as in Section
\ref{sec:E3} and that the calculation of the Ising energy of a
single state and the comparison with the minimum previously
obtained requires approximately $0.01$ seconds for a program on a
desktop computer. Then the total number of $2^{31}$ calculations
would already last longer than eight months. There exist
sophisticated methods to find Ising states which represent a
local, rather low energy minimum, see e.~g.~\cite{BP:2001}, but
for these methods we cannot be sure that we have found the global
minimum. In our case of
AB systems the following
simplification is possible: It suffices to check all Ising states
of a subsystem, say, $s_\mu,\;\mu\in{\mathcal A}$. If the
${\mathcal A}$-spins are fixed, the remaining spins
$s_\nu,\;\nu\in{\mathcal B}$ are calculated according to
\begin{equation}\label{C3}
s_\nu =-\text{sign} \left( \sum_{\mu\in{\mathcal A}} J_{\mu\nu} s_\mu \right)
\text{ for all } \nu\in{\mathcal B}
\;.
\end{equation}
The sign according to (\ref{C3}) minimizes the energy of the
interaction of the $\nu$-th Ising spin with its neighbors and
hence must be assumed for the total Ising ground state. This
simplification reduces in our above example $2^{31}$ calculations
to $2^{15}$ ones and thus the time for the total
calculation from months to minutes.\\

An alternative method to calculate the exact Ising ground state is
the so-called ``branch and bound" method, see \cite{KH:1978}.
According to this method the ground state problem is viewed as the
problem of finding a minimum (or maximum) of a function defined on
the binary tree given by the possible signs of the individual
Ising spins, such that good trial state energy or ``bound" for the
minimum is available. When scanning through the various
possibilities of the tree, one can ``cut" those branches of the
tree which will never reach the bound, even if the most optimistic
expectation for the remaining energies is adopted. This results in
a considerable reduction of the calculation time for finding the
absolute minimum of the energy, see \cite{KH:1978}.

%%%%%%%%%%%%%%%%%%%%%%%%%%%%%%%%%%%%%%%%%%%%%%%%%%%%%%%%%%%%%%%%%%%%%%%%%%%%%%%%%%%%%%%%%%%%%%%%%
%%%%%%%%%%%%%%%%%%%%%%%%%%%%%%%%%%%%%%%%%%%%%%%%%%%%%%%%%%%%%%%%%%%%%%%%%%%%%%%%%%%%%%%%%%%%%%%%%
\section{Generating generalized Bell inequalities\label{sec:B}}
%%%%%%%%%%%%%%%%%%%%%%%%%%%%%%%%%%%%%%%%%%%%%%%%%%%%%%%%%%%%%%%%%%%%%%%%%%%%%%%%%%%%%%%%%%%%%%%%%
\subsection{GBI from AB systems}\label{sec:AB}
%%%%%%%%%%%%%%%%%%%%%%%%%%%%%%%%%%%%%%%%%%%%%%%%%%%%%%%%%%%%%%%%%%%%%%%%%%%%%%%%%%%%%%%%%%%%%%%%%
\

In this Section we describe a recipe how to construct
new GBI's and provide a test whether they are
violated by the singlet state in QT. We proceed by
constructing frustrated spin systems.\\

\begin{itemize}
\item
We start by choosing an integer $N$ and a partition
of $\{1,\ldots,N\}$ into two disjoint subsets
${\mathcal A}$ and ${\mathcal B}$, not necessarily
with the same number of elements.
These sets correspond to the bipartition of the spin system
or to possible measurements performed by Alice and Bob.

\item Then we choose some real coefficients
$J_{\mu\nu},\;\mu\in{\mathcal A},\;\nu\in{\mathcal B}$.
They can be arbitrary but it is not advisable
to choose all coefficients with the same sign since we
are seeking for frustrated spin systems.

\item Next we find an Ising ground state, i.~e.~a sequence
$s_\mu=\pm 1, \; \mu=1,\ldots,N$ minimizing
the energy
$H^{(I)}=\sum_{\mu\in{\mathcal A}}\sum_{\nu\in{\mathcal B}}
J_{\mu\nu}\,s_\mu\, s_\nu$.
This can be done
by using the procedure described in the previous
Section \ref{sec:C}.
Note that any value $E$ of $H^{(I)}$ can be transformed into $-E$ by a spin
flip $s_\mu \mapsto -s_\mu$ for all $\mu \in \mathcal{A}$.
Let $-E_0^{(I)}$ denote the Ising
ground state energy. We thus obtain a GBI of the form
\begin{equation}\label{B1}
-E_0^{(I)} \le \sum_{\mu\in{\mathcal A}}
\sum_{\nu\in{\mathcal B}}J_{\mu\nu}\,
\langle s_\mu \,s_\nu \rangle \le E_0^{(I)}
\;.
\end{equation}
\item We perform a spin flip transformation
(\ref{E11a}),(\ref{E11b}) such that the Ising ground state becomes
$\uparrow\uparrow\ldots\uparrow$. Denote the transformed
coefficients again by $J_{\mu\nu}$.

\item We define a symmetric matrix $\mathbb{J}$ with
$J_{\mu\nu}=J_{\nu\mu}$ as non-diagonal elements.
The diagonal elements of $\mathbb{J}$ are chosen in such a way
that $\mathbb{J}$ will have constant row sums $j$ and vanishing
trace, see \cite{SM:2003}.
This leaves the Hamiltonian (\ref{G10}) unchanged.
Consequently, the Ising ground state $(1,1,\ldots,1)$ will be an
eigenvector of $\mathbb{J}$ with eigenvalue $j$.

\item We calculate the lowest eigenvalue
$j_{\text{\scriptsize min}}$ of
$\mathbb{J}$. If $j_{\text{\scriptsize min}}=j$ the Ising ground
state is already the Heisenberg ground state of the spin system
and our GBI will not be violated in QT. We have to start the
search anew. If, however, $j_{\text{\scriptsize min}}<j$ we are
done: We have found a GBI which is violated by the singlet state
in QT.
\end{itemize}

The single steps of this recipe are more or less obvious
except the last one. We know from (\ref{C1a}) that
$j_{\text{\scriptsize min}}<j$ is a necessary
condition for a spin system to have a ground state
energy $-E_0<-E_0^{(I)}$, but is it also sufficient? \\

In order to prove this we choose the fully polarized
Ising ground state to point into the $3$-direction
of our coordinate frame. Let $(x_\mu)_{\mu=1\ldots,N}$
denote an eigenstate of $\mathbb{J}$
with eigenvalue $j_{\text{\scriptsize min}}$.
Since, by assumption, $j_{\text{\scriptsize min}}<j$
this eigenstate is orthogonal to the Ising ground state,
i.~e.~$\sum_{\mu=1}^N x_\mu=0$.
Next we consider a smooth curve in the state space of
our spin system, i.~e.~a set of unit vector functions
$t\mapsto \vec{s}_\mu(t),\;\mu=1\ldots,N$ satisfying
\begin{equation}\label{B2}
\vec{s}_\mu(0)=
\left( \begin{array}{l}0\\0\\1\end{array}\right)
\text{  and    }
\frac{d}{d\,t}\vec{s}_\mu(0)=\left(
\begin{array}{l}x_\mu\\0\\0\end{array}
\right),\;\mu=1\ldots,N
\;.
\end{equation}
The Taylor expansion at $t=0$ of the Hamiltonian
evaluated at the states of this curve, $H(t)$, yields,
after a straightforward calculation,
using $\sum_{\mu=1}^N x_\mu=0$ and
$\frac{d^2}{d\,t^2}\vec{s}_\mu(0)\cdot\vec{s}_\mu(0)+
\frac{d}{d\,t}\vec{s}_\mu(0)\cdot\frac{d}{d\,t}
\vec{s}_\mu(0) =0$,
\begin{equation}\label{B3}
H(t)=H(0) + \frac{t^2}{2}(j_{\text{\scriptsize min}}-j)
\sum_{\mu=1}^N x_\mu^2 +{\mathcal O}(t^3)
\;.
\end{equation}
Hence $H(t)<H(0)$ for sufficiently small $t$ and
$-E_0^{(I)}$ cannot be the ground state energy.
This concludes the proof of the above statement.\\
Note, that we didn't prove that the energy
$\frac{N}{2}j_{\text{\scriptsize min}}$ is assumed by
some ground state. In general, this will not be the case.
The proof only shows that
$\frac{N}{2}j_{\text{\scriptsize min}}\le -E_0 <-E_0^{(I)}=\frac{N}{2}j$.

\subsection{Merging of spin sites}\label{sec:MSV}
The GBIs we have considered so far are generalizations of the CHSH inequality
involving spin measurements in the direction of four unit vectors $\vec{a}, \vec{b}, \vec{c}, \vec{d}$.
The original Bell inequality refers to measurements only belonging to three unit vectors and cannot
be associated with an AB system. To extend the connection between GBI’s and
frustrated spin systems to this and similar cases, it proves useful
to consider experimental tests of Bell’s inequality that are not based on spin measurements
but rather on measurements of photon pairs with perfect correlation of their linear
polarization. The unit vectors $\vec{a}, \vec{b}, \vec{c}, \ldots$ now refer to
the directions of the linear polarization.
We will recapitulate the derivation of the original Bell inequality,
adapted to the photon scenario. First, it should be noted that the CHSH inequality (\ref{G3})
also holds for photon experiments, since it was derived in a fairly general context.
Note that we may swap the two possible measurements $a$ and $c$ of Alice
and thus obtain a variant of the CHSH inequality of the form
\begin{equation}\label{G3a}
-2\le \langle c b \rangle +\langle c d\rangle +
\langle a b\rangle -\langle a d\rangle  \le 2
\;.
\end{equation}
We choose the direction $\vec{d}$ of Bob's polarization measurement so that it coincides with the
direction $\vec{c}$ of Alice's measurement. Furthermore, we assume perfect
correlation, meaning that for all possible outcomes, $d_i=c_i$ holds.
Upon the substitution $d\mapsto c$ which implies $\langle c d\rangle\mapsto +1$
the CHSH inequality (\ref{G3a}) entails
\begin{equation}\label{G3b}
\langle c b\rangle +1 +\langle a b\rangle-\langle a c \rangle \le 2
\end{equation}
and hence
\begin{equation}\label{G3c}
   \langle a b\rangle- \langle a c\rangle \le 1 -  \langle c b\rangle
   \;.
\end{equation}
Next we consider another variant of the CHSH inequality obtained
by the permutation $a\leftrightarrow c,\,b \leftrightarrow d $:
\begin{equation}\label{G3d}
-2\le \langle c d \rangle +\langle c b\rangle +
\langle a d\rangle -\langle a b\rangle  \le 2
\;.
\end{equation}
Upon the substitution $d\mapsto c$ we conclude
\begin{equation}\label{G3e}
1+\langle c b\rangle +1 +\langle a c\rangle-\langle a b \rangle \le 2
\end{equation}
and hence
\begin{equation}\label{G3f}
   \langle a c\rangle- \langle a b\rangle \le 1 -  \langle c b\rangle
   \;.
\end{equation}
Together with (\ref{G3c}) this implies
  \begin{equation}\label{G3f}
 \left|  \langle a c \rangle- \langle a b\rangle\right|\le 1 - \langle c b\rangle
   \;.
 \end{equation}
 The latter is the original Bell equation adapted to the photon scenario,
 see \cite{Bell:1964} and \cite{AP:1993}, eq.~(6.26),
 which involves only three polarization directions.

We now turn to the question of a possible violation of Bell's inequality
by photon pair experiments. We will only consider (\ref{G3c}) and its
quantum-theoretical counterpart
\begin{equation}\label{G3cq}
   \langle A B\rangle- \langle A C\rangle + \langle C B\rangle \le 1
   \;.
\end{equation}
The quantum-theoretical expression for the correlation $  \langle A B\rangle$
between linear polarization measurements with the directions $\vec{a}$ and $\vec{b}$ reads:
\begin{equation}\label{corrphoton}
  \langle A B\rangle = \cos 2 \measuredangle(\vec{a},\vec{b})
  \;,
\end{equation}
see \cite{AP:1993}, eq.~(6.16). The difference from the analog expression
(\ref{G6}) for the correlation between spin measurements lies, first, in the sign,
which reflects the difference between perfect correlation and anticorrelation, and
second, in the factor $2$ for the angle between the unit vectors.
For now, we will disregard the factor $2$ and only take it into account
again when designing an experiment that violates \ref{G3cq}).
We will thus write the inequality (\ref{G3cq}) in the form
\begin{equation}\label{G3dq}
 \vec{a} \cdot \vec{b} -  \vec{a} \cdot \vec{c} + \vec{c} \cdot \vec{b}  \le 1
\end{equation}
and interpret it as an upper bound of the energy of a classical spin triangle with
three bonds $+1,-1,+1$. The energy of a coplanar spin configuration
with $\measuredangle(\vec{a},\vec{b})=\measuredangle(\vec{b},\vec{c})=60^\circ$
and $\measuredangle(\vec{a},\vec{c})=120^\circ$ amounts to $E=3/2$ and hence violates (\ref{G3dq}).
Recall that the actual design of a photon pair experiment with linear polarization
along three unit vectors would have to be modified to
$\measuredangle(\vec{a}',\vec{b}')=\measuredangle(\vec{b}',\vec{c}')=30^\circ$
and $\measuredangle(\vec{a}',\vec{c}')=60^\circ$ due to the factor $2$ in (\ref{corrphoton}).

We observe that the violation of Bell's inequality concerns the \emph{upper}
bound of the energy of a classical spin system and has nothing to do with frustration.
Actually the spin triangle with bonds $+1,-1,+1$ is not frustrated. However, this
violation can be related with the frustrated spin triangle with inverted bonds, namely
$-1,+1,-1$, see Figure \ref{FIGT}. This possibility has to be taken into account for the
general problem of generating new GBI's by spin merging, see below.\\

Next we want to generalize the previous example and describe a procedure to
generate new GBI's from AB-systems.
Thus we start with a frustrated AB spin system with Hamiltonian
\begin{equation}\label{M1}
H=\sum_{\mu\in{\mathcal A}} \sum_{\nu\in{\mathcal B}}
J_{\mu\nu} \vec{s}_\mu\cdot\vec{s}_\nu
\;,
\end{equation}
Then we merge certain spins according to an equivalence relation $\gamma$
on ${\mathcal A} \cup {\mathcal B}$. We define the new set of spin sites
${\mathcal C}$ by the set of equivalence classes
\begin{equation}\label{defC}
  {\mathcal C} := \left( {\mathcal A} \cup {\mathcal B}\right)/_\gamma
\end{equation}
and new coupling constants $\widetilde{J}_{\alpha\beta}$ by
\begin{equation}\label{defJtilde}
 \widetilde{J}_{\alpha\beta}:= \sum_{\mu\in\alpha,\,\nu\in\beta} J_{\mu\,\nu}
 \quad
 \mbox{for all } \alpha\neq\beta\in  {\mathcal C}
 \;.
\end{equation}
The two new spin Hamiltonians can be defined by
\begin{equation}\label{newH}
 \widetilde{H}^\pm =\pm\,\sum_{\alpha,\beta\in{\mathcal C}}  \widetilde{J}_{\alpha\beta}
 \vec{s}_\alpha\cdot\vec{s}_\beta
 \;,
\end{equation}
where the contribution due for the interaction between spins of the same equivalence class
gives only a constant energy shift and can be neglected.
The two  new spin systems (\ref{newH}) are candidates for a frustrated system that
has a coplanar ground state energy below the Ising ground state energy and hence
leads to a new GBI. However, this is generally not certain.
Let’s consider the Bell square as a counterexample
and merge two adjacent spins at a time. The resulting spin system is a dimer,
which is never frustrated.

While the AB systems considered in Subsection \ref{sec:AB} are, at most,
non-geometrically frustrated, the new systems arising from spin merging
could exhibit a kind of combination of geometric and non-geometric frustration,
as the previous example shows. In this respect, these systems resemble spin glasses
in which the couplings are randomly distributed. However, it seems highly unlikely
that general spin glasses can be regarded as the product of AB systems following spin merging.\\

%%%%%%%%%%%%%%%%%%%%%%%%%%%%%%%%%%%%%%%%%%%%%%%%%%%%%%%%%%%%%%%%%%%%%%%%%%%%%%%%%%%%%%%%%%%%%%%%%
%%%%%%%%%%%%%%%%%%%%%%%%%%%%%%%%%%%%%%%%%%%%%%%%%%%%%%%%%%%%%%%%%%%%%%%%%%%%%%%%%%%%%%%%%%%%%%%%%
\section{Examples\label{sec:E}}
%%%%%%%%%%%%%%%%%%%%%%%%%%%%%%%%%%%%%%%%%%%%%%%%%%%%%%%%%%%%%%%%%%%%%%%%%%%%%%%%%%%%%%%%%%%%%%%%%
%%%%%%%%%%%%%%%%%%%%%%%%%%%%%%%%%%%%%%%%%%%%%%%%%%%%%%%%%%%%%%%%%%%%%%%%%%%%%%%%%%%%%%%%%%%%%%%%%

According to Section \ref{sec:B} arbitrary many
GBI's can be constructed. Nevertheless,
it will be instructive to show how known
examples of GBI's fit into our scheme and to
consider further examples for which the
ground states can be calculated in closed form.

%%%%%%%%%%%%%%%%%%%%%%%%%%%%%%%%%%%%%%%%%%%%%%%%%%%%%%%%%%%%%%%%%%%%%%%%%%%%%%%%%%%%%%%%%%%%%%%%%
%%%%%%%%%%%%%%%%%%%%%%%%%%%%%%%%%%%%%%%%%%%%%%%%%%%%%%%%%%%%%%%%%%%%%%%%%%%%%%%%%%%%%%%%%%%%%%%%%
\subsection{The frustrated $2n$-ring\label{sec:E1}}
%%%%%%%%%%%%%%%%%%%%%%%%%%%%%%%%%%%%%%%%%%%%%%%%%%%%%%%%%%%%%%%%%%%%%%%%%%%%%%%%%%%%%%%%%%%%%%%%%
%%%%%%%%%%%%%%%%%%%%%%%%%%%%%%%%%%%%%%%%%%%%%%%%%%%%%%%%%%%%%%%%%%%%%%%%%%%%%%%%%%%%%%%%%%%%%%%%%

As a generalization of the frustrated Bell square
with $N=4$ we consider a spin ring with an even
number $N=2n$ of spins, cyclic boundary condition
$N+1\equiv 1$ and nearest neighbor interaction $J_{\mu,\mu+1}=\pm 1,\,\mu=0,\ldots,N-1$.
This is an AB system since the couplings exists only between even and odd spin sites.
We will shortly summarize the results of \cite{FSJ:2026} as they
pertain to this specific spin ring.

This ring is only frustrated in the case of an odd number of
anti-ferromagnetic bonds $J_{\mu,\mu+1}=1$.
Its energy is unchanged by a spin flip transformation
\begin{eqnarray}\label{E11a}
\vec{s}_\mu &\mapsto& \delta_\mu \vec{s}_\mu\\ \label{E11b}
J_{\mu\nu} &\mapsto& \delta_\mu \delta_\nu J_{\mu\nu}
\;,
\end{eqnarray}
where $ \mu,\nu=0,\ldots,N-1$ and $\delta_\mu=\pm 1$.
Hence the transformed ground state will be the
ground state of the transformed Hamiltonian.
Using this spin flip transformation a spin ring with
an odd number of anti-ferromagnetic bonds can be transformed
into a spin ring with only ferromagnetic bonds except a single
antiferromagnetic one, say, $J_{01}=1$.

For the moment we will consider the slightly more
general case of a frustrated spin polygon with $N=2n$ sites and ferromagnetic bonds
$J_{\mu,\mu+1}=-1$ except one single anti-ferromagnetic bond of strength $J_{01}=\alpha_0>0$.
Then it follows from \cite{FSJ:2026}, theorem 2, that the polygon will have
a collinear ground state for $0< \alpha_0 \le \frac{1}{N-1}$ and a coplanar one, at least,
for $\frac{1}{N-1}\le \alpha_0 \le 1$. For the coplanar ground state in the case $\alpha_0=1$
it follows from \cite{FSJ:2026}, prop.~1, that all angles $\psi$ between spin vectors with ferromagnetic bonds
$J_{\mu,\mu+1}=-1,\, \mu=1, 0<\mu<N$
are equal to $\psi=\pi/N$ and that the ground state energy will be
\begin{equation}\label{E0}
  -E_0 = - N \, \cos \frac{\pi}{N}
  \;.
\end{equation}
This result also holds in the case of odd $N$.
For $N\ge 3$ the ground state energy (\ref{E0}) is strictly lower than the Ising ground state energy $-E_0^{(I)}=-N+2$
realized by the fully polarized state $\uparrow\uparrow\ldots \uparrow$.

The GBI corresponding to frustrated $2n$-rings has been found by
P.~M.~Pearle \cite{P:1970}, eq.~(27),
and its violation by quantum correlations
has been discussed by S.~L.~Braunstein and C.~M.~Caves
\cite{BC:1990}, but without proving the maximal
violation for the configuration described above.

%%%%%%%%%%%%%%%%%%%%%%%%%%%%%%%%%%%%%%%%%%%%%%%%%%%%%%%%%%%%%%%%%%%%%%%%%%%%%%%%%%%%%%%%%%%%%%%%%
%%%%%%%%%%%%%%%%%%%%%%%%%%%%%%%%%%%%%%%%%%%%%%%%%%%%%%%%%%%%%%%%%%%%%%%%%%%%%%%%%%%%%%%%%%%%%%%%%
\subsection{The frustrated hexagon\label{sec:E2}}
%%%%%%%%%%%%%%%%%%%%%%%%%%%%%%%%%%%%%%%%%%%%%%%%%%%%%%%%%%%%%%%%%%%%%%%%%%%%%%%%%%%%%%%%%%%%%%%%%
%%%%%%%%%%%%%%%%%%%%%%%%%%%%%%%%%%%%%%%%%%%%%%%%%%%%%%%%%%%%%%%%%%%%%%%%%%%%%%%%%%%%%%%%%%%%%%%%%

%===================    figure   =================================
\begin{figure}[ht]
\begin{center}
\includegraphics[width=80mm]{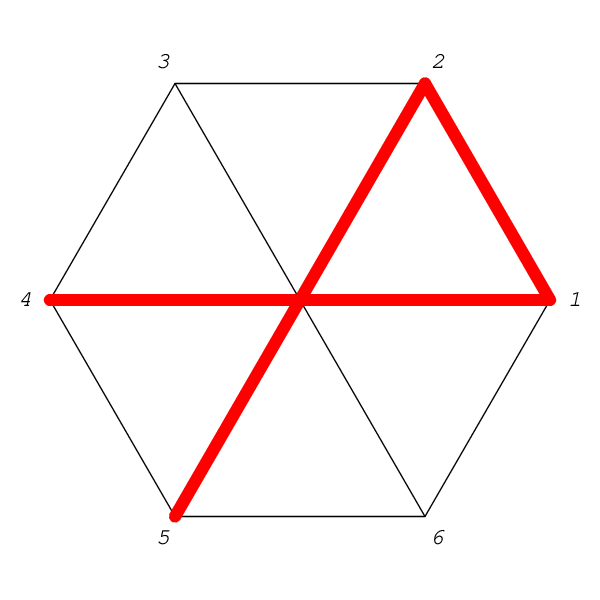}
\caption{\label{fig01}A frustrated hexagon. The thick red lines indicate the edges with a
coupling constant of $-1$, the thin black lines correspond to the coupling $+1$.
}
\end{center}
\end{figure}

%===================    figure   =================================
\begin{figure}[ht]
\begin{center}
\includegraphics[width=80mm]{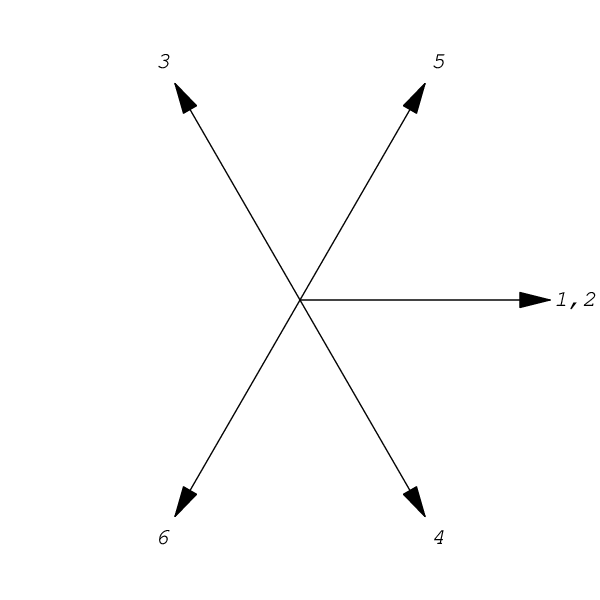}
\caption{\label{fig02}A coplanar ground state of the frustrated hexagon.
The correlations of measurements of the
spin components according to these vectors in the singlet state violate
the inequality (\ref{E22}).
}
\end{center}
\end{figure}

Frustrated even rings are not the only AB-systems. We consider as another example
the frustrated hexagon with additional interactions
between even and odd spin sites, see figure \ref{fig01}.
Its coupling constants are chosen as follows:
\begin{eqnarray}\label{E21a}
J_{12}
&=&
J_{14}=J_{25}=-1\\ \label{E21b}
J_{23}
&=&
J_{34}=J_{45}=J_{56}=J_{16}=J_{36}=1
\;.
\end{eqnarray}
The corresponding GBI reads
\begin{equation}\label{E22}
-5\le \langle s_2 s_3\rangle + \langle s_3 s_4\rangle + \langle
s_4 s_5\rangle + \langle s_5 s_6\rangle + \langle s_6 s_1\rangle +\langle
s_3 s_6\rangle -\langle s_1 s_2\rangle -\langle s_1 s_4\rangle
-\langle s_2 s_5\rangle \le 5 \;.
\end{equation}
corresponding to the Ising ground state
$\uparrow\uparrow\uparrow\downarrow\uparrow\downarrow$.
The spin configuration with
\begin{equation}\label{E23}
\vec{s}_1=\vec{s}_2,\;
\sphericalangle(\vec{s}_6,\vec{s}_4)=
\sphericalangle(\vec{s}_4,\vec{s}_2)=
\sphericalangle(\vec{s}_2,\vec{s}_5)=
\sphericalangle(\vec{s}_5,\vec{s}_3)=60^\circ
\;,
\end{equation}
see figure \ref{fig02}, realizes the Heisenberg ground state with
energy $-E_0=-6$ corresponding to a violation of (\ref{E22}) in
QT. Since this result can be confirmed by using the same methods
as in Subsection \ref{sec:E1}, we leave the details to the reader.

%%%%%%%%%%%%%%%%%%%%%%%%%%%%%%%%%%%%%%%%%%%%%%%%%%%%%%%%%%%%%%%%%%%%%%%%%%%%%%%%%%%%%%%%%%%%%%%%%
%%%%%%%%%%%%%%%%%%%%%%%%%%%%%%%%%%%%%%%%%%%%%%%%%%%%%%%%%%%%%%%%%%%%%%%%%%%%%%%%%%%%%%%%%%%%%%%%%
\subsection{Frustrated hypercubes\label{sec:E3}}
%%%%%%%%%%%%%%%%%%%%%%%%%%%%%%%%%%%%%%%%%%%%%%%%%%%%%%%%%%%%%%%%%%%%%%%%%%%%%%%%%%%%%%%%%%%%%%%%%
%%%%%%%%%%%%%%%%%%%%%%%%%%%%%%%%%%%%%%%%%%%%%%%%%%%%%%%%%%%%%%%%%%%%%%%%%%%%%%%%%%%%%%%%%%%%%%%%%
\subsubsection{Definition, general properties and low-dimensional cases \label{sec:DGP}}
%%%%%%%%%%%%%%%%%%%%%%%%%%%%%%%%%%%%%%%%%%%%%%%%%%%%%%%%%%%%%%%%%%%%%%%%%%%%%%%%%%%%%%%%%%%%%%%%%

\begin{table}
\caption{\label{tab1}Results for the ground states of some frustrated hypercubes $H_n,\;n=2,3,4,5$.
The spin vectors $\vec{s}_\mu$ of the ground state are only given for $\mu=0,2,\ldots,N-2$; the other
spin vectors result from the symmetry (\ref{E33}). Moreover, we utilize the abbreviations
(\ref{E34a}), (\ref{E34b}), (\ref{E34c}), and (\ref{E34d}).
The Ising ground states are degenerate; the table contains only one example of an Ising ground state
for each $n=2,3,4,5$.
We observe that for $n=2,3,5$ the
Heisenberg ground state energy $-E_0^{(n)}$ is below the energy $-E_0^{(n,I)}$ of the Ising ground state.
This corresponds to a violation of the generalized Bell inequalities in quantum theory.
For $n=4$ both ground state energies coincide, hence the corresponding
generalized Bell inequality is \emph{not} violated in quantum theory.
}
\vspace{5mm}
\item[]\begin{tabular}{@{}llp{65mm}p{30mm}ll}
\hline
$n$ & $N=2^n$ &  Heisenberg ground state & $-E_0^{(n)}$ & Ising ground state & $-E_0^{(n,I)}$\\
\hline
$2$
&
$4$
&
$\vec{s}_\mu=V_2(\delta_1,\delta_2,\delta_3)$, where $(\delta_1,\delta_2,\delta_3)=$
&
$-2\sqrt{2}\approx -2.82843$
&
$\uparrow \downarrow\downarrow\uparrow $
&
$-2$\\
&& $(+ + -),(- - +)$
&&&\\
\hline $3$ & $8$ &
$\vec{s}_\mu=V_3(\delta_1,\delta_2,\delta_3,\delta_4)$, where
$(\delta_1,\delta_2,\delta_3,\delta_4)=$ & $-4\sqrt{3}\approx
-6.9282$ &
$\uparrow\uparrow\downarrow\downarrow\downarrow\downarrow\downarrow\uparrow$
&
$-6$\\
&& $(+ - + -),(- + - +),
(- - + +),(- - - -)$
&&&\\
\hline
$4$
&
$16$
&
$\vec{s}_\mu=V_4(\delta_1,\delta_2)$, where $(\delta_1,\delta_2)=$
&
$-8\sqrt{4}=-16$
&
$\downarrow\downarrow\uparrow\uparrow\uparrow\uparrow\downarrow\uparrow\uparrow
\uparrow\uparrow\downarrow\downarrow\uparrow\uparrow\downarrow$
&
$-16$\\
&& $(-1 0),(+1 0),(+1 0),(0 -1),$\newline $(+1 0),(0 +1),(0 -1),(0 +1)$
&&&\\
\hline
$5$
&
$32$
&
$\vec{s}_\mu=V_5(\delta_1,\delta_2,\delta_3)$, where $(\delta_1,\delta_2,\delta_3)=$
&
$-16\sqrt{5}\approx -35.7771$
&
$\uparrow\uparrow\downarrow\downarrow\downarrow
\downarrow\downarrow\downarrow\downarrow\downarrow
\uparrow\uparrow\downarrow\downarrow\uparrow\uparrow
$
\newline
$\uparrow\uparrow\uparrow\uparrow\uparrow
\uparrow\uparrow\uparrow\uparrow\uparrow
\downarrow\downarrow\uparrow\uparrow\uparrow\uparrow
$
&
$-32$\\
&&
(+, --, -1), (--, --, 2),
(+, +, 2),  (+, +, -2), \newline
(+, +, -1),  (--, +, -1),
(--, +, -1),  (--, --, 2),  \newline
(--, +, 1), (+, +, -2),
(--, --, -2),  (+, +, 2),\newline
(--, --, 1), (+, --, 1),
(+, --, 1), (+, +, -2)
&&&\\
\hline\hline
\end{tabular}
\end{table}
%-------------------------table-------------------------------

%%%%%%%%%%%%%%%%%%%%%%%%%%%%%%%%%%%%%%%%%%%%%%%%%%%%%%%%%%%%%%%%%%%%%%%%%%%%%%%%%%%%%%%%%%%%%%%%%
Another way to generalize the Bell square is to
consider ``frustrated hypercubes" $H_n$ with
$N=2^n$ vertices and $n=1,2,3,\ldots$.
For other applications of hypercubes in connection with spin systems
see  \cite{MC:2015} and \cite{CMNS:2025}.
In our case, the $N$
spins are located at the vertices and are
interacting along the $n 2^{n-1}$ edges of the
hypercube.
Sometimes, the vertices $v_\nu,\;\nu=0,\ldots,N-1$
are most conveniently labelled by binary strings
$(\delta_1,\ldots,\delta_n)$
of length $n$, such that the $\delta_i\in\{0,1\}$
are the binary digits of $\nu$.
Two strings $v,w$
which differ exactly at one position form an edge
$e=\{v,w\}$.
The set of vertices can be divided into the set
${\mathcal A}$ of binary strings with
an even number of ones and the set ${\mathcal B}$
of binary strings with
an odd number of ones. Thus every edge connects an
${\mathcal A}$-vertex with
a ${\mathcal B}$-vertex and the corresponding
spin system is an AB-system.
The sign attached to an edge $e=\{v,w\}$
can be defined as $(-1)^\ell$ where $\ell$ is
the number of ones at the left hand side
of the position where $v$ and $w$ differ.\\

Alternatively, the signs can be
defined recursively according to the following
procedure.\\
If the signs of $H_n$ are already defined, denote by ${H_n^\ast}$
the same hypercube with inverted signs. Define $H_{n+1}$ as the
union of $H_n$ and ${H_n^\ast}$ where $N$ new edges between the
corresponding vertices of $H_n$ and ${H_n^\ast}$ are
added and $+1$-signs are attached to them.\\
For example, if we start with $H_1=$
\unitlength1mm\begin{picture}(17,6)
\put(2,1.5){\circle{3}}\put(14,1.5)
{\circle{3}}\put(3.5,1.5){\line(1,0){9}}\put(7,3){+}
\end{picture}
and apply this procedure we obtain
\vspace{-1cm}
\begin{equation}\label{E31}
H_2=
\unitlength1mm\begin{picture}(20,20)(-4,7)
\put(2,1.5){\circle{3}}\put(14,1.5){\circle{3}}
\put(3.5,1.5){\line(1,0){9}}\put(7,-1){--}
\put(2,13.5){\circle{3}}\put(14,13.5){\circle{3}}
\put(3.5,13.5){\line(1,0){9}}\put(7,15){+}
\put(2,3){\line(0,1){8.5}}\put(-1,6.5){+}
\put(14,3){\line(0,1){8.5}}\put(14,6.5){+}
\end{picture}
\end{equation}
\vspace{5mm}
\\
which is just the Bell square considered before. See figure
\ref{fig03} for the frustrated hypercubes $H_3$ and $H_4$ obtained
by this procedure. Of course, one can apply symmetry
transformations of $H_n$ (suitable rotations and reflections) and
spin flip transformations to obtain modified frustrated
hypercubes. But these will have analogous properties as the $H_n$
and need not be considered further.\\
The recursive procedure $H_n\rightarrow
H_{n+1}$ can also be applied to the $\mathbb{J}$-matrices.
Indeed, $\mathbb{J}_n$ can be recursively defined by
\begin{eqnarray}\label{E32a}
\mathbb{J}_0
&=&
(0)\\ \label{E32b}
\mathbb{J}_{n+1}
&=&
\left(\begin{array}{rr}\mathbb{J}_{n} & \mathbb{I}_{N}\\
\mathbb{I}_{N}&-\mathbb{J}_{n}\end{array}
\right)
\;,
\end{eqnarray}
where $\mathbb{I}_{N}$ denotes the $N\times N$
identity matrix.
By induction we conclude $\mathbb{J}_{n}^2=
n \mathbb{I}_{N}$ and hence the eigenvalues of
$\mathbb{J}_{n}$ are $\pm\sqrt{n}$ with
degeneracies $\frac{N}{2}$. If these eigenvalues
can be realized by spin vector configurations
the corresponding ground state energy
$-E_0^{(n)}=-\frac{N}{2}\sqrt{n}$ must be lower
than the Ising ground state energy $-E_0^{(n,I)}$
which is always an integer. Possible exceptions
are $n=4,9,16,\ldots$ where $E_0^{(n)}$ is an integer
too.\\

We cannot present a general theorem which warrants that the ground
state energy $-E_0^{(n)}=-\frac{N}{2}\sqrt{n}$ will be assumed by
a coplanar state for all $n\ge 2$. Rather we have studied the
cases $n=3,4,5$ in some detail with the following results.\\
For $n=3$ and $n=5$ we have explicitly constructed coplanar spin
configurations realizing the ground state energy
$-E_0^{(n)}=-\frac{N}{2}\sqrt{n}$ which is below the Ising ground
state energy. In these cases we therefore obtain GBI's which are
violated in QT. For $n=4$ the ground state energy $-E_0^{(4)}=-16$
is realized by an Ising state. Hence we can also in this case
write down a GBI, but it is also satisfied in QT and hence of no
use in the context of the EPR discussion. This GBI is also not
violated for other entangles states, see Section \ref{sec:S}.\\

The frustrated hypercubes $H_n$ have an obvious reflection
symmetry $\sigma$ defined by
$\nu\stackrel{\sigma}{\longleftrightarrow} \nu+1, \nu \text{
even}$. We thus have $J_{\sigma(\mu)\sigma(\nu)}=J_{\mu\nu}$. This
can by proven by induction over $n$ using the recursive
construction procedure $H_n\rightarrow H_{n+1}$. Hence the
eigenvectors of $\mathbb{J}$ can be chosen either to be invariant
under $\sigma$ or to change their sign. Recall that the $x$- and
the $y$-components of the coplanar spin configuration realizing
the ground state energy $-E_0^{(n)}=-\frac{N}{2}\sqrt{n}$ are
eigenvectors of $\mathbb{J}$. It turns out that we can always find
ground state configurations
\begin{equation}\label{E33}
\vec{s}_\mu={x_\mu \choose y_\mu} \text{  satisfying   }
x_{\sigma(\mu)}=x_\mu\text{ and }y_{\sigma(\mu)}=-y_\mu \text{ for
}{\mu=1,\ldots,N} \;.
\end{equation}
Hence it will suffice to write down the ground state
configurations $\vec{s}_\mu$ only for even $\mu$.
\\

All components of $\vec{s}_\mu$ can be written as radicals
with a similar structure
consisting of nested square roots and small integers.
In order to simplify table \ref{tab1} containing the
ground state configurations we hence introduce
the following abbreviations:
\begin{eqnarray}\label{E34a}
n=2&:&
V_2(\delta_1,\delta_2,\delta_3)
\equiv
\frac{1}{2}\left(
\begin{array}{l}\delta_1\sqrt{2+\delta_3 \sqrt{2}}\\
\delta_2\sqrt{2-\delta_3 \sqrt{2}}
\end{array}
\right)
\\ \label{E34b}
n=3&:&
V_3(\delta_1,\delta_2,\delta_3,\delta_4)
\equiv
\frac{1}{2}\left(
\begin{array}{l}\delta_1\sqrt{2+\delta_3 \sqrt{\frac{2}{3}}+\delta_4\frac{2}{\sqrt{3}}}\\
\delta_2\sqrt{2-\delta_3 \sqrt{\frac{2}{3}}-\delta_4\frac{2}{\sqrt{3}}}
\end{array}
\right)
\\ \label{E34c}
n=4&:&
V_4(\delta_1,\delta_2)
\equiv
\left(
\begin{array}{l}\delta_1\\
\delta_2
\end{array}
\right)
\\ \label{E34d}
n=5&:&
V_5(\delta_1,\delta_2,\delta_3)
\equiv
\left(
\begin{array}{l}\delta_1\sqrt{\frac{1}{2}+\frac{1}{\delta_3 \sqrt{5}}}\\
\delta_2\sqrt{\frac{1}{2}-\frac{1}{\delta_3 \sqrt{5}}}
\end{array}
\right)
\;.
\end{eqnarray}

\begin{figure}
\begin{center}
\includegraphics[width=100mm]{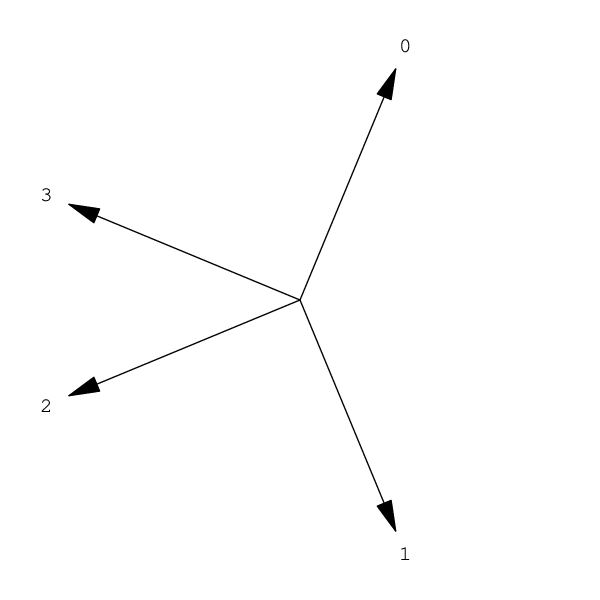}
\caption{\label{fig04a}A coplanar ground state of the frustrated
Bell square $H_2$ according to table \ref{tab1}. Note the reflection symmetry of the
vectors with label $\nu$ and $\nu+1$, $\nu$ even.
The correlations of measurements of the spin components according to these vectors
in the singlet state violate the CHSH inequality.
}
\end{center}
\end{figure}

\begin{figure}
\begin{center}
\includegraphics[width=100mm]{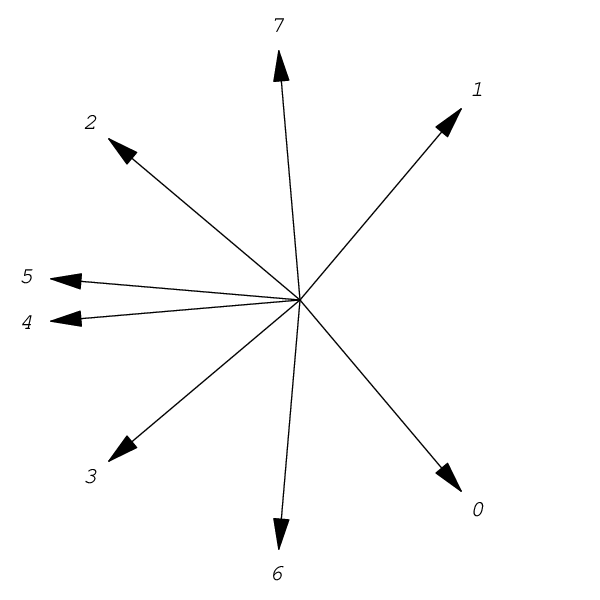}
\caption{\label{fig04}A coplanar ground state of the frustrated
cube $H_3$ according to table \ref{tab1}. Note the reflection symmetry of the
vectors with label $\nu$ and $\nu+1$, $\nu$ even.
The correlations of measurements of the spin components according to these vectors
in the singlet state violate the corresponding generalized Bell inequality.
}
\end{center}
\end{figure}

\begin{figure}
\begin{center}
\includegraphics[width=100mm]{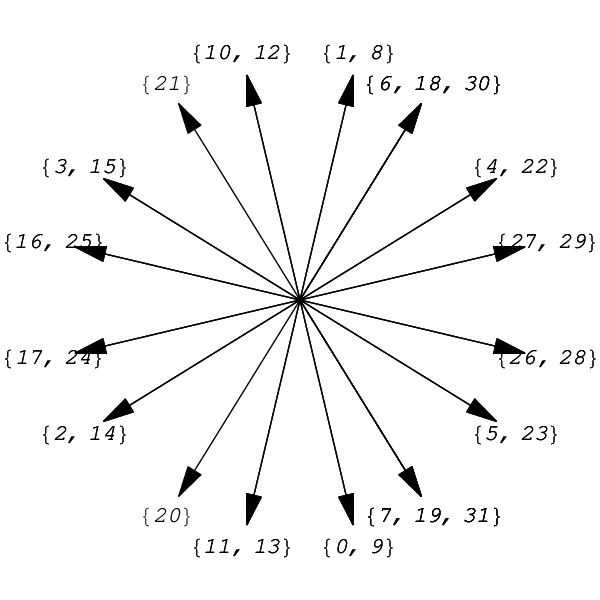}
\caption{\label{fig05}A coplanar ground state of the frustrated
hypercube $H_5$ according to table \ref{tab1}. Note the reflection symmetry of the
vectors with label $\nu$ and $\nu+1$, $\nu$ even.
The correlations of measurements of the spin components according to these vectors
in the singlet state violate the corresponding generalized Bell inequality.
}
\end{center}
\end{figure}

The coplanar ground states for some frustrated hypercubes $H_n$
are visualized in the figures \ref{fig04a} ($n=2$), \ref{fig04} ($n=3$) and
\ref{fig05} ($n=5$).

%===================    figure   =================================
%===================    figure   =================================
%===================    figure   =================================
%===================    figure   =================================

%%%%%%%%%%%%%%%%%%%%%%%%%%%%%%%%%%%%%%%%%%%%%%%%%%%%%%%%%%%%%%%%%%%%%%%%%%%%%%%%%%%%%%%%%%%%%%%%%
\subsubsection{Symmetries of frustrated hypercubes}\label{sec:SFH}
%%%%%%%%%%%%%%%%%%%%%%%%%%%%%%%%%%%%%%%%%%%%%%%%%%%%%%%%%%%%%%%%%%%%%%%%%%%%%%%%%%%%%%%%%%%%%%%%%

As noted in Section \ref{sec:DGP}, we cannot prove in general
that the frustrated hypercubes $H_n$ have coplanar ground states with energy $-E_0= \frac{N}{2\sqrt{n}}$.
We have only been able to confirm this for the cases $n=2,3,4,5$.
In this Section, however, we will present some additional arguments
to support this conjecture, partly because doing so will help us
understand the structure of $H_n$ and its ground states 

We consider the general case of a classical spin system with the Hamiltonian
\begin{equation}\label{genHam}
 H(\mathbf{s}) = \sum_{\mu\nu} J_{\mu\nu}\, \vec{s}_\mu \cdot \vec{s}_\nu\;.
 \end{equation}
Generally, the ground state(s) cannot be constructed by linear combinations
of the eigenvectors corresponding to the lowest eigenvalue of the coupling matrix $J$.
However, in many cases this can be achieved if the matrix $J$ is replaced by
the ``dressed $J$-matrix" $\widetilde{J}({\boldsymbol \lambda})$ defined by
\begin{equation}\label{defdJ}
 \widetilde{J}({\boldsymbol \lambda})_{\mu\nu}= J_{\mu\nu}\, +\lambda_\mu\,\delta_{\mu\nu},
 \quad \quad \mu,\nu=1,\ldots N
 \;,
\end{equation}
such that
\begin{equation}\label{condlambda}
  \sum_\mu \lambda_\mu =0
  \;,
\end{equation}
see \cite{S:2017} and \cite{SR:2022}.

The meaning of $\lambda_\mu$ follows from the equation
\begin{equation}\label{SSE}
 \sum_\nu J_{\mu\nu}\,\vec{s}_\nu = -\kappa_\mu \,\vec{s}_\mu, \quad\quad \mu=1,\ldots, N
 \;,
\end{equation}
which results from minimizing $H({\mathbf s})$ subject to the constraints
$\| \vec{s}_\mu\|^2 = 1$ for $\mu = 1, \ldots, N$ with the corresponding
Lagrange multipliers $\kappa_\mu \ge 0$.
Using the definitions $\overline{\kappa}:= \frac{1}{N}\sum_{\mu=1}^N \kappa_\mu$
and $\lambda_\mu:= \kappa_\mu-\overline{\kappa}$  we obtain
from (\ref{SSE}) the eigenvalue equation
\begin{equation}\label{eigJd}
 \sum_\nu \underbrace{\left( J_{\mu\nu}+\lambda_\mu\,\delta_{\mu\nu}\right)}_
 {=\widetilde{J}({\boldsymbol \lambda})_{\mu\nu}}
 \,\vec{s}_\nu = - \overline{\kappa}\,\vec{s}_\mu, \quad\quad \mu=1,\ldots, N
\end{equation}
for the dressed $J$-matrix.
The above conjecture is thus essentially equivalent to the statement
$\lambda_\mu=0$ for all $\mu=1,\ldots,N$, i.~e., to the constancy of the
Lagrange multipliers
(there is a complication due to the dimensionality of the ground state, more on this later).
In many cases, the constancy of the Lagrange multipliers is a consequence
of the high symmetry of the $J$ matrix. This holds, for example,
for Bravais spin lattices, to which the Luttinger–Tisza theory of
ground states applies; see \cite{SR:2022}.

However, for the frustrated hypercubes $H_n$, we have so far found
only a single permutation symmetry $\sigma$ that swaps even and odd spin sites.
On the other hand, the fact that the two eigenvalues
$\pm \sqrt{n}$ of $J_n$ have the  maximal degeneracy of $N/2$ is an indication of high symmetry.
We will therefore extend the concept of permutation symmetry and
define “signed permutations,” which are represented by real
$N\times N$ matrices $L$ with
\begin{equation}\label{signper}
 \sum_{\nu=1}^N L_{\mu\nu}\, \vec{s}_\nu = \delta_\mu\, \vec{s}_{\pi(\nu)} =: {\vec{r}_\mu},
 \quad \mbox{ for all } \mu=1,\ldots, N
 \;,
\end{equation}
where $\left| \delta_\mu\right|=1$ and $\pi\in{\mathcal S}_N$ denotes a permutation.

As an example consider
\begin{equation}\label{exsignper}
 \left(
 \begin{array}{ccc}
   0 & 1 & 0 \\
   0 & 0 & 1 \\
   -1 & 0 & 0
 \end{array}
 \right) \,\left(
 \begin{array}{c}
   \vec{s}_1 \\
   \vec{s}_2 \\
   \vec{s}_3
 \end{array}
 \right)=
 \left(
 \begin{array}{c}
   \vec{s}_2 \\
   \vec{s}_3 \\
   -\vec{s}_1
 \end{array}
 \right)
 \;.
\end{equation}
This signed permutation matrix commutes with the $J$-matrix of the triangle
according to Figure \ref{FIGT}:
\begin{equation}\label{LJJL}
 \left(
 \begin{array}{ccc}
   0 & 1 & 0 \\
   0 & 0 & 1 \\
   -1 & 0 & 0
 \end{array}
 \right) \,
 \left(
 \begin{array}{ccc}
   0 & -1 & 1 \\
   -1 & 0 & -1 \\
   1 & -1 & 0
 \end{array}
 \right)
 =
  \left(
 \begin{array}{ccc}
   0 & -1 & 1 \\
   -1 & 0 & -1 \\
   1 & -1 & 0
 \end{array}
 \right) \,
  \left(
 \begin{array}{ccc}
   0 & 1 & 0 \\
   0 & 0 & 1 \\
   -1 & 0 & 0
 \end{array}
 \right)
 =  \left(
 \begin{array}{ccc}
   -1 & 0 & -1 \\
   1 & -1 & 0 \\
   0 & 1 & -1
 \end{array}
 \right)
 \;.
\end{equation}

With this definition the following Propositiom can be formulated:\\

\begin{prop}\label{PS}
Let $\left( \vec{s}_\mu\right)_{\mu=1,\ldots, N}$ be a classical ground state
of a spin system with Hamiltonian
\begin{equation}\label{genHam1}
  H(\mathbf{s}) = \sum_{\mu\nu} J_{\mu\nu}\, \vec{s}_\mu \cdot \vec{s}_\nu
  \;,
\end{equation}
with a symmetric coupling matrix $J$ satisfying
\begin{equation}\label{SSE1}
 \sum_\nu J_{\mu\nu}\,\vec{s}_\nu = -\kappa_\mu \,\vec{s}_\mu, \quad\quad  \mbox{for all } \mu=1,\ldots, N
 \;.
\end{equation}
Further, let $L$ be a signed permutation matrix, i.~e.~,
\begin{equation}\label{signper1}
 \sum_{\nu=1}^N L_{\mu\nu}\, \vec{s}_\nu = \delta_\mu\, \vec{s}_{\pi(\nu)} =: \vec{r}_\mu,
 \quad \mbox{ for all } \mu=1,\ldots, N
 \;,
\end{equation}
such that $J \,L=L\,J$.
Then $\left( \vec{r}_\mu\right)_{\mu=1,\ldots, N}$ will again be a ground state of (\ref{genHam1})
satisfying
\begin{equation}\label{SSE2}
 \sum_\nu J_{\mu\nu}\,\vec{r}_\nu = -\kappa'_\mu \,\vec{r}_\mu, \quad\quad \mu=1,\ldots, N
 \;,
\end{equation}
where
\begin{equation}\label{kappap}
 \kappa'_\mu = \kappa_{\pi(\mu)} \quad \quad \mbox{for all } \mu=1,\ldots,N
 \;.
\end{equation}
\end{prop}

We will omit the straightforward proof here.
It is important to note that, due to a signed permutation, which
is a symmetry of the spin system, the Lagrangian multipliers of a
ground state are only permuted by $\pi$
(a change in sign would also be fatal because $\kappa_\mu \ge 0$).

It can be proven that, in the general case, the Lagrange multiplier $\kappa_\mu,\,\mu=1,\ldots, N$
have the same values for all ground states of a spin system, see theorem 3 in \cite{S:2017}.
It follows, in conjunction with Proposition \ref{PS}, that the $N$-tuple
$(\kappa_1,\ldots, \kappa_N)$ is invariant under $\pi$, even in the case of a signed permutation.
The resulting equations $\kappa_\mu=\kappa_{\pi(\mu)}$ restrict the number of independent Lagrange multipliers.
In the extreme case of maximum symmetry, only one independent value
$\kappa_\mu=\overline{\kappa}$ remains, and the dressed $J$-matrix with ${\boldsymbol \lambda}=\mathbf{0}$
can be replaced by the original $J$-matrix.

We return to the case of the frustrated hypercubes and must therefore
examine $J_n$ for further symmetries of the ``signed permutation" type.
We have found such symmetries, which, interestingly, can be expressed as
generators of an $s=1/2$ representation of the group $SU(2)$,
not to be confused with the usual $SO(3)$ symmetry of the Hamiltonian (\ref{genHam}).
The Lie algebra $su(2)$ will be realized by certain anti-Hermitean matrices of the form
\begin{equation}\label{defLn}
L_{n} = \left(
\begin{array}{cc}
  A_n(\alpha) &  B_n(\beta,\gamma) \\
  B_n(\beta,\gamma) &  A_n(\alpha)
\end{array}
\right),\quad \alpha,\beta,\gamma \in {\mathbbm R}
\;,
\end{equation}
which are recursively defined as follows:
\begin{equation}\label{AnBn}
 A_n(\alpha)= \left(
\begin{array}{cc}
  A_{n-1}(\alpha) &  {\mathbf 0} \\
 {\mathbf 0} &  A_{n-1}(\alpha)
\end{array}
\right), \mbox{ and }
B_n(\beta,\gamma)=\left(
\begin{array}{cc}
  B_n(\beta,\gamma) &  {\mathbf 0} \\
 {\mathbf 0} &  - B_n(\beta,\gamma)
\end{array}
\right)
\;,
\end{equation}
starting with
\begin{equation}\label{A2B2}
A_2(\alpha) =\frac{1}{2}\left(\begin{array}{cc}
                                    0 & -{\sf i} \alpha \\
                                    -{\sf i} \alpha & 0
                                  \end{array} \right),\quad
B_2(\beta,\gamma) =\frac{1}{2}\left(\begin{array}{cc}
                                   -{\sf i}\gamma & \beta \\
                                    -\beta& {\sf i}\gamma
                                  \end{array} \right)
\;.
\end{equation}

It can be shown by induction that the $L_n$ are symmetries of the frustrated hypercubes $H_n$, i.~e.~,
\begin{equation}\label{Lnsym}
 L_n\,J_n = J_n\, L_n,\quad \mbox{ for all } n=2,3,\ldots
\end{equation}

Due to the recursive structure (\ref{E32b}) of the $J_n$,
further symmetries  arise from the $L_n$ which are of the form
\begin{equation}\label{L1n}
 L(1,n):= \left(
 \begin{array}{cc}
   L_{n-1} & {\mathbf 0} \\
  {\mathbf 0} &  L_{n-1}
 \end{array}
 \right) = {\mathbb I}_2\otimes L_{n-1}
 \;,
\end{equation}
\begin{equation}\label{L2n}
 L(2,n):= \left(
 \begin{array}{cccc}
   L_{n-2} & {\mathbf 0} & {\mathbf 0} & {\mathbf 0} \\
  {\mathbf 0} & L_{n-2} & {\mathbf 0} & {\mathbf 0} \\
   {\mathbf 0} &{\mathbf 0} & L_{n-2} &{\mathbf 0}\\
   {\mathbf 0} & {\mathbf 0} & {\mathbf 0} & L_{n-2}
 \end{array}
 \right) = {\mathbb I}_4\otimes L_{n-2}
 \;,
\end{equation}
and, generally,
\begin{equation}\label{Lnun}
 L(\nu,n):= {\mathbb I}_{2^\nu}\otimes L_{n-\nu}\quad \mbox{for } \nu=1,2,\ldots, n-2
 \;,
\end{equation}

If $(\alpha,\beta,\gamma)$ is a standard unit vector,
these symmetries are, up to an irrelevant factor, of the type ``signed permutation".
Based on the above considerations, the $N$-tuple $(\kappa_1,\ldots,\kappa_N)$
must then be invariant under the corresponding permutations.
In the cases  $n=2,3,4,5$ examined, these conditions are sufficient
to ensure constant Lagrange multipliers $\kappa_\mu\equiv \overline{\kappa}$
for the ground states.
It is easy to prove that this must hold for general $n\ge 2$
since each symmetry $L(\nu,n)$ adds another permutation under which
the Lagrange multiplier $(\kappa_1,\ldots,\kappa_N)$ must be invariant.
This shows that the classical ground states
of the frustrated hypercubes can be constructed from the eigenvectors
corresponding to the lowest eigenvalue $-\sqrt{n}$.

However, this does not necessarily imply that these ground states
are always coplanar (or, in specific cases such as $n=4$, even collinear).
Also $3$-dimensional ground states would be acceptable.
But there are cases where the ground states constructed using the method
described above — which is explained in more detail in \cite{S:2017} — have a
non-physical dimension $d>3$. The physical ground states would then
have an energy above $-N/2 \sqrt{n}$, and violation of the GBI would not be guaranteed.
Although this possibility exists, we suspect that it does not occur in the case of frustrated hypercubes.

%%%%%%%%%%%%%%%%%%%%%%%%%%%%%%%%%%%%%%%%%%%%%%%%%%%%%%%%%%%%%%%%%%%%%%%%%%%%%%%%%%%%%%%%%%%%%%%%%
%%%%%%%%%%%%%%%%%%%%%%%%%%%%%%%%%%%%%%%%%%%%%%%%%%%%%%%%%%%%%%%%%%%%%%%%%%%%%%%%%%%%%%%%%%%%%%%%%
\section{General entangled states\label{sec:S}}
%%%%%%%%%%%%%%%%%%%%%%%%%%%%%%%%%%%%%%%%%%%%%%%%%%%%%%%%%%%%%%%%%%%%%%%%%%%%%%%%%%%%%%%%%%%%%%%%%
%%%%%%%%%%%%%%%%%%%%%%%%%%%%%%%%%%%%%%%%%%%%%%%%%%%%%%%%%%%%%%%%%%%%%%%%%%%%%%%%%%%%%%%%%%%%%%%%%
In the correspondence between GBI's and frustrated AB-systems
outlined in Section \ref{sec:G} we have always chosen
the singlet state
$\phi=\frac{1}{\sqrt{2}}\left( |\,\uparrow\downarrow\,\rangle -
|\,\downarrow\uparrow\,\rangle\right)$
as the state relative to which the correlations
$\langle AB \rangle$ etc. have been calculated.
The rotational invariance of $\phi$ then directly
corresponds to the rotational invariance
of the Heisenberg Hamiltonian (\ref{G10}).
This picture changes if other entangled states
$\psi\in{\mathcal H}\equiv \mathbb{C}^2\otimes \mathbb{C}^2$
are taken into account.\\

Such states can be written as a Schmidt bi-orthogonal sum
\begin{equation}\label{S1}
\psi=\sum_{i=1}^2 c_i \mathbf{u}_i\otimes\mathbf{v}_i
\;,
\end{equation}
where $\{\mathbf{u}_i\}_{i=1,2}$ and $\{\mathbf{v}_i\}_{i=1,2}$
are orthonormal bases in $\mathbb{C}^2$,
and $|c_1|^2+|c_2|^2=1$, see \cite{AP:1993}, Section 5.~3.
$\psi$ is an entangled state if and only if $c_1, c_2\neq 0$.
Upon choosing appropriate, possibly different,
coordinate frames for Alice and Bob, (\ref{S1}) assumes
the form
\begin{equation}\label{S2}
\psi=\cos\alpha |\uparrow\downarrow\rangle-\sin\alpha |\downarrow\uparrow\rangle
\;,
\end{equation}
where $0<\alpha <\frac{\pi}{2}$.
$\alpha=\frac{\pi}{4}$ corresponds to the singlet state $\phi$.
To put this in another way, $\psi$ can be written in the form
(\ref{S2}) by means of a unitary
transformation in ${\mathcal H}$.\\

After a straightforward calculation we obtain for the correlation of
observables of the form (\ref{G5})
\begin{eqnarray}\label{S3a}
\langle AB \rangle
&\equiv&
\langle\psi | A\otimes B |\psi\rangle\\ \label{S3b}
&=&
-a_3 b_3-\sin(2\alpha) (a_1 b_1 +a_2 b_2)\\ \label{S3c}
&\equiv&
-a_3 b_3-\gamma (a_1 b_1 +a_2 b_2)
\;.
\end{eqnarray}
The corresponding classical spin Hamiltonian for a general
frustrated AB-system reads
\begin{equation}\label{S4}
H(\gamma)=\sum_{\mu\in{\mathcal A}}\sum_{\nu\in{\mathcal B}} J_{\mu\nu}
\left(z_\mu z_\nu +\gamma (x_\mu x_\nu + y_\mu y_\nu )\right)
\;,
\end{equation}
where we have written
\begin{equation}\label{S5}
\vec{s}_\mu\equiv \left(\begin{array}{l}x_\mu\\y_\mu\\z_\mu\end{array}\right)
\text{ for }\mu=1,\ldots,N
\;.
\end{equation}
Such anisotropic spin Hamiltonians are well-known under the name
``XXZ-model", see, for example \cite{FB:2004}. For
$\gamma\rightarrow\infty$ the XXZ-model essentially approaches the
Ising model, which means that, in the context of GBI's,
factorable states will not violate the GBI.\\
It has been proven \cite{GP:1992},\cite{AP:1993} that the CHSH
inequality is also violated for an arbitrary entangled state
$\psi$, even for a general Hilbert space ${\mathcal H}$. This
means, in the language of spin systems, that the XXZ Bell square
has a coplanar ground state for all $\gamma$ with $0<\gamma\le 1$.
Unfortunately, we have not found a proof of the analogous
statement for the class of GBI's considered in this article. We only
remark that, obviously, the GBI is violated for all $\gamma$ with
$E_0^{(I)}/E_0<\gamma<1$, if $E_0^{(I)}<E_0$
and the GBI is violated for $\gamma=1$ with a coplanar ground
state. Indeed, if we evaluate the Hamiltonian $H(\gamma)$ at the
coplanar ground state $(\vec{s}_\mu)_{\mu=1,\ldots,N}$ we obtain
\begin{eqnarray}\label{S6a}
H(\gamma)
&=&
\sum_{\mu\in{\mathcal A}}\sum_{\nu\in{\mathcal B}} J_{\mu\nu}
(\underbrace{z_\mu z_\nu}_{=0} +\gamma (x_\mu x_\nu + y_\mu y_\nu ))
\\ \label{S6b}
&=&
\gamma H(1) =-\gamma E_0< - E_0^{(I)}
\;.
\end{eqnarray}

We now consider the case that the Ising ground state is an
eigenstate of $\mathbb{J}$
corresponding to the lowest eigenvalue
$j_{\text{\scriptsize min}}$, and hence
the GBI is {\emph not} violated for the singlet state $\phi$.
This happens
in the example of the frustrated hypercube $H_4$, see table \ref{tab1}.
Then the GBI will
also not be violated for other entangled states $\psi$, i.~e.~,
the Ising ground state will
remain the ground state for all $H(\gamma)$.
To show this we consider
\begin{eqnarray}\label{S7a}
2H(\gamma)
&=&
\sum_{\mu,\nu=1}^N J_{\mu\nu}
(z_\mu z_\nu+\gamma (x_\mu x_\nu + y_\mu y_\nu ))\\ \label{S7b}
&\ge&
j_{\text{\scriptsize min}}(
\underbrace{\sum_{\mu=1}^N z_\mu^2}_{\zeta} +\gamma
\underbrace{\sum_{\mu=1}^N(x_\mu^2 +y_\mu^2)}_{\xi}) \\ \label{S7c}
&\equiv&
j_{\text{\scriptsize min}}(\zeta +\gamma\xi)\ge N j_{\text{\scriptsize min}}
\;,
\end{eqnarray}
since $j_{\text{\scriptsize min}}<0$ and
$\zeta +\gamma\xi$ assumes its maximum under the constraint
$\zeta +\xi=N$ for $\zeta=N$.
The minimum (\ref{S7c}) is assumed by inserting the Ising
ground state
for the $z_\mu$ and setting $x_\mu=y_\mu=0$.\\

%%%%%%%%%%%%%%%%%%%%%%%%%%%%%%%%%%%%%%%%%%%%%%%%%%%%%%%%%%%%%%%%%%%%%%%%%%%%%%%%%%%%%%%%%%%%%%%%%
%%%%%%%%%%%%%%%%%%%%%%%%%%%%%%%%%%%%%%%%%%%%%%%%%%%%%%%%%%%%%%%%%%%%%%%%%%%%%%%%%%%%%%%%%%%%%%%%%
\section{Conclusion\label{sec:CC}}
%%%%%%%%%%%%%%%%%%%%%%%%%%%%%%%%%%%%%%%%%%%%%%%%%%%%%%%%%%%%%%%%%%%%%%%%%%%%%%%%%%%%%%%%%%%%%%%%%
%%%%%%%%%%%%%%%%%%%%%%%%%%%%%%%%%%%%%%%%%%%%%%%%%%%%%%%%%%%%%%%%%%%%%%%%%%%%%%%%%%%%%%%%%%%%%%%%%
What are the benefits of the proposed correspondence
between the considered class of GBI's and
frustrated AB-systems?\\

For readers who are mainly interested in the EPR discussion and
Bell inequalities the most interesting result might be the recipe
to obtain an arbitrary number of GBI's, see Section \ref{sec:B}.
This construction procedure should give us some additional insight
into the structure of GBI's. Moreover, we hope that the proposed
correspondence would lead the reader to think of GBI's in a more
geometric or graphical way: The possible measurements can be
visualized as vertices of a spin system and the signed
correlations occurring in the GBI as the edges or bonds of this
system. As we have shown, some methods from the theory of spin
systems can be employed to tackle questions in the realm of the
foundations of QT.\\

Also for readers which are mainly interested in frustrated spin
systems, the unexpected connection to GBI's might be valuable in its
own right. For problems in the theory of spin systems a transfer
of methods from another field of research will be of some
interest. Moreover, those readers will probably find the class of
frustrated hypercubes, Subsection \ref{sec:E3}, to be a useful set of
toy examples worth while to be further studied. For example, in
the quantum XY-model of frustrated hypercubes there exist
so-called localized multi-magnon states which lead to prominent
properties as huge magnetization jumps and large residual entropy
at $T=0$ resulting in a marked magneto-caloric effect, see, for
example \cite{RSH:2004} and \cite{JS:2004}. \\

Frustrated spin systems with an AB structure are by no means exotic constructs;
on the contrary, there are examples of them that have been studied
in detail in the literature, such as spin ladders with mixed F/AF interactions.
These can be ladders with one F leg and another AF leg, see \cite{MDK:2018} \cite{CKS:2024},
or ladders with AF-leg interactions and F-rung interactions,
see \cite{H:1995}. A material with the latter property is, for example,
\ce{Rb Fe2 Se3}, see \cite{Wetal:2016}.
\\

Summarizing, the correspondence between GBI's and frustrated spin
systems fosters the insight into both branches of physics and
leads to some new phenomena in non-geometric frustration.

%%%%%%%%%%%%%%%%%%%%%%%%%%%%%%%%%%%%%%%%%%%%%%%%%%%%%%%%%%%%%%%%%%%%%%%%%%%%%%%%%%%%%%%%%%%%%%%%%
%%%%%%%%%%%%%%%%%%%%%%%%%%%%%%%%%%%%%%%%%%%%%%%%%%%%%%%%%%%%%%%%%%%%%%%%%%%%%%%%%%%%%%%%%%%%%%%%%
%\section*{Acknowledgement}
%%%%%%%%%%%%%%%%%%%%%%%%%%%%%%%%%%%%%%%%%%%%%%%%%%%%%%%%%%%%%%%%%%%%%%%%%%%%%%%%%%%%%%%%%%%%%%%%%
%%%%%%%%%%%%%%%%%%%%%%%%%%%%%%%%%%%%%%%%%%%%%%%%%%%%%%%%%%%%%%%%%%%%%%%%%%%%%%%%%%%%%%%%%%%%%%%%%

%I thank Klaus B\"arwinkel, Marshall Luban, Johannes Richter, and J\"urgen Schnack
%for a critical reading of the manuscript.


\begin{thebibliography}{99}


\bibitem{Bell:1964}
J.~S. Bell, On the Einstein Podolski Rosen paradox, Physics
\textbf{1}, 195--200 (1964).


\bibitem{EPR:1935}
A. Einstein, B. Podolski, and N. Rosen,
Can quantum-mechanical description be
considered complete? Phys. Rev. \textbf{47}, 777--780
(1935).




\bibitem{BJ:1987}
L.~E.~Ballantine and J.~P.~Jarrett, Bell's theorem:
Does quantum mechanics contradict
relativity?, Am. J. Phys. \textbf{55} (8), 696--701 (1987)



\bibitem{CHSH:1969}
J.~F. Clauser, M.~A. Horne, A. Shimony, and R.~A. Holt,
Proposed experiment to test local hidden-variable theories,
Phys. Rev. Lett. \textbf{23}, 880--884 (1969).

\bibitem{P:1970}
P.~M.~Pearle, Hidden-variable example based upon data rejection,
Phys. Rev. D \textbf{8}, 1418--1425 (1970)



\bibitem{BC:1990}
S.~L. Braunstein and C.~M. Caves,
Wringing out better Bell inequalities, Ann. Phys. (NY)
\textbf{202}, 22--56 (1990).

\bibitem{GS:1979}
A.~Garuccio and F.~Selleri, Systematic derivation of all inequalities of Einstein locality,
Found. Phys.\textbf{10}, 209--216, (1979)

\bibitem{AP:1993}
A. Peres, \emph{Quantum Theory: Concepts and Methods},
Kluwer (1993).



\bibitem{EHAT:2024}
P.~Emonts, M.~Hu, A.~Aloy, and J.~Tura,
Effects of topological boundary conditions on Bell nonlocality,
\textit{Phys. Rev. A}, {\bf 110} (3), 032201 (2024)


\bibitem{Hetal:2026}
M.~Hu, E.~Vall{\'e}e, T.~Seynnaeve, P.~Emonts, F.~Mohammadi, and J.~Tura,
Characterizing translation-invariant Bell inequalities using tropical algebra and graph polytopes,
\textit{Phys. Rev. A}, {\textbf 113} (3), 032421 (2026)


\bibitem{AP:1978}
A.~Peres, Unperformed experiments have no results,
\textit{Am. J. Phys.} \textbf{46}, 745--747 (1978)

\bibitem{MR:2006}
R.~Moessner and A.~Ramirez, Geometrical Frustration,
\textit{Physics Today}, {\textbf 59} (2), 24--29 (2006)

\bibitem{FSJ:2026}
W.~Florek, H.-J.~Schmidt, and K.~Ja{\'s}niewicz-Pacer,
Ground states of classical spin polygons: rigorous results and examples
\textit{ Z. Naturforsch. A}, {\textbf 81} (4), 243--263 (2026)


\bibitem{SM:2003}
H.~-~J.~Schmidt and M.~Luban, Classical ground states
of symmetric Heisenberg spin systems,
J. Phys. A:~Math.~Gen.~\textbf{36}, 63514--6378 (2003)

\bibitem{SR:2022}
H.~-~J.~Schmidt and J.~Richter, Classical ground states of spin lattices,
J. Phys. A:~Math.~Gen.~\textbf{55}, (46)  465005 (2022)


\bibitem{BP:2001}
S.~Boettcher and A.~G.~Percus, Optimization with extremal dynamics,
Phys. Rev. Lett. \textbf{86}, 5211--5214 (2001)


\bibitem{KH:1978}
S.~Kobe and A.~Hartwig, Exact ground states of finite amorphous Ising systems,
Comp. Phys. Commun. \textbf{16} 1, 1--4  (1978)

\bibitem{GP:1992}
N.~Gisin, A.~Peres,
Maximal violation of Bell inequality for arbitrarily large spin,
Phys.Letters A  \textbf{162}, 15--17 (1992)

\bibitem{FB:2004}
D.~J.~J.~Farnell and R.~F.~Bishop, chapter 7 in \cite{SRFB:2004}

\bibitem{RSH:2004}
J.~Richter, J.~Schulenburg, and A.~Honecker,
chapter 2 in \cite{SRFB:2004}

\bibitem{JS:2004}
J.~Schnack, chapter 3 in \cite{SRFB:2004}

\bibitem{SRFB:2004}
U.~Schollw\"ock, J.~Richter, D.~J.~J.~Farnell, and R.~F.~Bishop (Eds.~),
\emph{Quantum Magnetism}, Lecture Notes in Physics 645, Springer (2004)

\bibitem{MC:2015}
A.~Mozeika  and A.~C.~Coolen,  Spin systems on hypercubic Bethe lattices: a Bethe–Peierls approach.
\textit{J. Phys. A: Math. Theor.}, {\bf 48} (25), 25500 (2015)

\bibitem{CMNS:2025}
R.~Chen, J. Machta, C.~M.~Newman, and D.~L.~Stein,
Zero-temperature dynamics of Ising systems on hypercubes,
\textit{Phys. Rev. E }, {\bf 112} (5), 054135 (2025)

\bibitem{S:2017}
H.-J.~Schmidt,
Theory of ground states of classical Heisenberg spin systems I,
Preprint,
arXiv:1701.02489v2,\; (2017)


\bibitem{SR:2022}
H.-J.~Schmidt and J.~Richter,  Classical ground states of spin lattices.
\textit{J. Phys. A: Math. Theor.}, {\bf 55} (46), 465005 (2022)


\bibitem{MDK:2018}
D.~Maiti, D.~Dey, and M.~Kumar, Frustrated spin-1/2 ladder with ferro-and antiferromagnetic legs.
\textit{JMMM}, {\bf 446} , 170--176 (2018)


\bibitem{CKS:2024}
M.~Chatterjee, M.~Kumar, and Z.~G.~Soos, Singlet quantum phases of the frustrated spin-1/2
ladder with ferromagnetic (F) exchange in legs and alternating F-AF exchange in rungs.
\textit{Phys. Scr.}, {\bf 99}  (2), 025973 (2024)


\bibitem{H:1995}
K.~Hida, Density matrix renormalization group study of the spin 1/2 Heisenberg
ladder with antiferromagnetic legs and ferromagnetic rungs.
\textit{J. Phys. Soc. Jpn.}, {\bf 64} (12), 170--176 (2018)


\bibitem{Wetal:2016}
M.Wang et al,  Spin waves and magnetic exchange interactions in the spin-ladder compound \ce{Rb Fe2 Se3}.
\textit{Phys. Rev. B}, {\bf 94} (4), 4896--4900 (1995)


\end{thebibliography}
\end{document}